\documentclass[journal=jacsat,manuscript=article]{achemso}
\usepackage{chemformula} % Formula subscripts using \ch{}
\usepackage[T1]{fontenc} % Use modern font encodings
\usepackage{multirow} % for table cells to span rows
\usepackage{amsfonts}
\usepackage{amsmath} % matrix
\usepackage{amssymb}
\usepackage{mathtools}
\usepackage{epsfig}
\usepackage{braket}
\usepackage{float}
\usepackage{graphicx}
\usepackage[version=4]{mhchem}
\usepackage{array}
\usepackage[columnwise]{lineno}
\usepackage{threeparttable}
\usepackage{enumitem}
\usepackage{caption}
\usepackage{url}
\usepackage{mciteplus}
\mciteErrorOnUnknownfalse % gets rid of false reference errors
\usepackage[nolist]{acronym}
\usepackage[frozencache,cachedir=.,outputdir=.]{minted} % use on local machine before submission
\usepackage{xcolor} % to access the named colour LightGray
\usepackage{rotating}
\usepackage{csquotes}
\usepackage{hyperref}
\hypersetup{hidelinks}
\newcommand{\dftb}{\textsc{DFTB+}}
\newcommand{\mopac}{\textsc{MOPAC}}
\newcommand{\fhiaims}{\textsc{FHI-aims}}

\newcommand{\lammps}{\textsc{LAMMPS}}
\newcommand{\mosdef}{\textsc{MoSDeF}}
\newcommand{\qca}{\textsc{QCArchive}}
\newcommand{\gaussian}{\textsc{Gaussian}}
\newcommand{\bibtex}{\textsc{Bib}\TeX}
\newcommand{\psifour}{\textsc{Psi4}}
\newcommand{\etal}{\emph{et~al.}}
\newcommand{\etc}{\emph{etc.}}

\newcommand{\eg}{\emph{e.g.}}
\newcommand{\vv}{\emph{vice versa}}
\newcommand{\defacto}{\emph{de facto}}
\newcommand{\si}{\text{Supporting Information}}
\newcommand{\abinit}{\emph{ab initio}}
\newcommand{\Eh}{E_\text{h}}
\newcommand{\kcalmol}{kcal mol$^{-1}$}

\newcommand{\wn}{\text{cm}^{-1}}
\newcommand{\gcmcub}{\text{g cm}^{-3}}
\newcommand{\mr}{\multirow}

\author{Paul Saxe}
\email{psaxe@vt.edu}
\affiliation{Department of Chemistry, Virginia Tech, Blacksburg, Virginia 24061, USA}
\alsoaffiliation{Molecular Sciences Software Institute, Blacksburg, Virginia 24060, USA}
\author{Jessica Nash}
\email{janash@vt.edu}
\affiliation{Department of Chemistry, Virginia Tech, Blacksburg, Virginia 24061, USA}
\alsoaffiliation{Molecular Sciences Software Institute, Blacksburg, Virginia 24060, USA}
\author{Mohammad Mostafanejad}
\email{smostafanejad@vt.edu}
\affiliation{Department of Chemistry, Virginia Tech, Blacksburg, Virginia 24061, USA}
\alsoaffiliation{Molecular Sciences Software Institute, Blacksburg, Virginia 24060, USA}
\author{Eliseo Marin-Rimoldi}
\email{emarinri@nd.edu}
\affiliation{Department of Chemical and Biomolecular Engineering, University of Notre Dame,\
Notre Dame, Indiana 46556, USA}
\author{Hasnain Hafiz}
\email{hasnain.hafiz@gm.com}
\affiliation{Battery Research and Development, General Motors, Warren, MI 48092, USA}
\author{Louis G Hector, Jr.}
\email{louis.hector@gm.com}
\affiliation{Battery Research and Development, General Motors, Warren, MI 48092, USA}
\author{T. Daniel Crawford}
\email{crawdad@vt.edu}
\affiliation{Department of Chemistry, Virginia Tech, Blacksburg, Virginia 24061, USA}
\alsoaffiliation{Molecular Sciences Software Institute, Blacksburg, Virginia 24060, USA}
\title{SEAMM: A Simulation Environment for Atomistic and Molecular Modeling}

\begin{document}

% \begin{tocentry}

  % Some journals require a graphical entry for the Table of Contents.
  % This should be laid out ``print ready'' so that the sizing of the

  % Inside the \texttt{tocentry} environment, the font used is Helvetica
  % 8\,pt, as required by \emph{Journal of the American Chemical
  %   Society}.

  % The surrounding frame is 9\,cm by 3.5\,cm, which is the maximum
  % permitted for  \emph{Journal of the American Chemical Society}
  % graphical table of content entries. The box will not resize if the
  % content is too big: instead it will overflow the edge of the box.

  % This box and the associated title will always be printed on a
  % separate page at the end of the document.

% \end{tocentry}

% Do not use acronyms operators \ac{} in the abstract to facilitate copy-pasting to journal forms
\begin{abstract}
The Simulation Environment for Atomistic and Molecular Modeling (SEAMM)  is an open-source software package written in Python that provides a graphical interface for setting up, executing, and analyzing molecular and materials simulations. The graphical interface reduces the entry barrier for the use of new simulation tools, facilitating the interoperability of a wide range of simulation tools available to solve complex scientific and engineering problems in computational molecular science. Workflows are represented graphically by user-friendly flowcharts which are shareable and reproducible. When a flowchart is executed within the SEAMM environment, all results, as well as metadata describing the workflow and codes used, are saved in a datastore that can be viewed using a browser-based dashboard, which allows collaborators to view the results and use the flowcharts to extend the results. SEAMM is a powerful productivity and collaboration tool that enables interoperability between simulation codes and ensures reproducibility and transparency in scientific research.

\end{abstract}

\section{Introduction}
\ac{CMS} and computational materials science (also \ac{CMS} -- we will use the term to encompass both) have been growing in importance since their beginning over 70 years ago. Figure 1 shows the growth of citations of the major \ac{CMS} software packages in the international scientific literature as well as international patents, which are proxies for basic and applied research.
\begin{figure}[!tbph]
	\centering
	\includegraphics{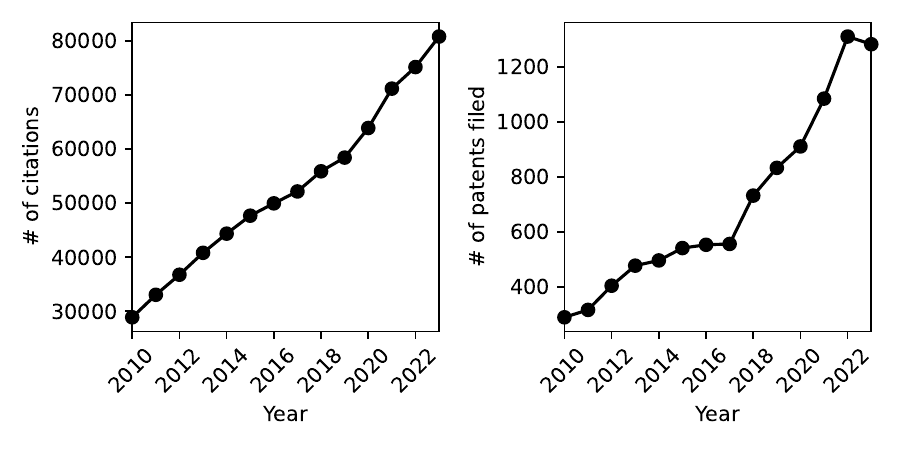}
	\caption{Number of citations\cite{Talirz:2021:1483,Talirz:2024:zenodo.10894860} (left panel) and patents (right panel) filed using significant CMS codes.}
	\label{FIG:CitPat}
\end{figure}Citation statistics for the \ac{CMS} codes, obtained from the \texttt{atomistic.software} website,\cite{Talirz:2021:1483,Talirz:2024:zenodo.10894860} are based on the Google Scholar tracking engine. The number of patents is also drawn from Google Patents. Similar trends are reported by other research groups,\cite{Haunschild:2016:1,Dumaz:2021:6681} adopting bibliometric analysis of subjects such as \ac{DFT}, which is the theory underlying a significant part of the \ac{CMS} simulations. 

The growth of annual citations of \ac{CMS} codes is exponential with an approximate doubling time of about 9.5 years. The number of patents citing \ac{CMS} codes is much smaller ($\approx$1,300 vs. 80,000 in 2013) but the growth is faster (doubling almost every 6 years). This is faster than the overall scientific publishing counts, which is growing at about 4-5\% per year with a doubling time of about 15 years.\cite{bornmann2021growth}
Clearly, both the development and application of \ac{CMS} codes is of significant and increasing importance.

The growth of \ac{CMS} as a field is mainly driven by two major factors. The technological factor, pertaining to hardware and the incredible increase in the available computing power, demonstrates a recent doubling time of about 20 months.\cite{top500} The computational factor, on the other hand, focuses on the continual development of the underlying theory and practical implementations in software. Optimal control in both factors allows researchers to perform more accurate calculations on a larger scale that cover a wider range of practical use cases.

How are \ac{CMS} codes used in practice? Individual codes are often adopted to simulate a physical system, e.g. a molecule or collection of molecules, a fluid, a crystalline or an amorphous solid, in order to predict its structure and properties, such as energy, band gap, or spectra. These simulations have many use cases. They can be used to explore systems of interest, to gain a better understanding, to help interpret and guide experiments, or to predict various properties, which often inform downstream or secondary student models. In some cases, a single code and type of simulation is sufficient to 
obtain a preset threshold of accuracy for simple and small systems. However, often numerous simulations, performed using different codes, are required to study and analyze the properties of larger and more complex systems. For example, optimizing the components of a battery might require quantum chemical simulations to determine reduction potentials, molecular dynamics simulations to predict diffusion constants and ionic conductivity of the electrolyte, and vibrational thermodynamics from phonon calculations to predict the entropy coefficient of anode and cathode materials for the reversible volumetric heat source term in battery thermal models.

It usually takes tens or hundreds of simulations to understand the system, discover the hidden patterns within a measured set of properties and optimize these properties under certain conditions for practical purposes. Data management is also a significant challenge for simulations, as it requires careful tracking and documentation of the tools, methods, and parameters used. Furthermore, data management practices directly affect the reproducibility of the generated results and the extensibility of the performed simulations to new molecules or materials. At the end of the scholarly research process, the downstream tasks such as writing reports, manuscripts, or patents summarizing the results of the experiments will also be affected by the data management practices.

Some projects are handled by a single researcher, but others may require a team of researchers with computational and experimental backgrounds in different parts of the problem. Some researchers might handle the simulations, while others use the results to interpret and assist experiments or as input to other models that operate at different length and time scales. This collaboration increases the need for careful organization and tracking of the artifacts of the simulations.

However, \ac{CMS} codes are not easy to use: the underlying theory is complex and the resulting equations are difficult to solve exactly except for the simplest "toy" systems, leading to many different approximations to the exact equations. Quantum chemistry codes have a range of methods from Hartree-Fock and \ac{DFT}, which have larger approximations but are less computationally expensive, to coupled-cluster methods and full-CI, which are computationally very expensive but are closer to the exact solution. In \ac{DFT}, the exact functional is not known, so there are numerous approximate functionals available. Most quantum codes make another approximation by expanding the wavefunction in terms of basis functions, which are typically Gaussian functions for molecules or plane waves for periodic systems. There are hundreds of different Gaussian basis sets in use.\cite{bse} \acp{MD} codes are also complicated with many choices of forcefields or atomic potentials, many different algorithms to simulate different ensembles, and a wide variety of analyses. This complexity is unavoidable in any code trying to unravel chemistry and physics at the atomic level.\cite{doi:10.1021/jacs.2c13042}

This complexity is obvious in the input files that the user must prepare to run a simulation. There are a daunting number of input parameters to choose from. Some combinations of parameters may be allowed and others not, and if one parameter is set to a certain value, others may need to be adjusted to be compatible. Major quantum chemistry codes such as \gaussian\cite{Pople:2003:GAUSSIAN03,g16} and GAMESS\cite{GAMESS} have hundreds up to a few thousand keywords to choose from. \ac{VASP}\cite{vasp1,vasp2,vasp3,vasp4}, an electronic structure code widely used in the materials field, has about 46 main keywords, many of which have suboptions, yielding hundreds of combinations. This complexity is not restricted to quantum calculations: \iac{LAMMPS}\cite{lammps1,lammps2}, an \ac{MD} code that is also widely used in the solid-state field, has over 1,000 commands, many of which have suboptions. It takes time and effort (and considerable trial and error) to learn how to use one of these simulation codes correctly and effectively. Unfortunately, while experience with one code lowers the barrier to learning a similar code, it only reduces it a moderate amount because the input and output of different codes are quite different. Experience with one class of codes is very little help in learning to use a different type of code. Many expert users of quantum chemistry codes are not comfortable with \ac{MD} codes, and \vv. Similarly, users of solid-state codes such as \ac{VASP} face a considerable barrier with quantum chemistry codes, and \vv.

The same issues make it difficult to compare codes or to reproduce or replicate the work of other researchers. It is challenging to include enough detail, even in the supplementary information, to be able to reproduce the results using the same code. It is considerably more difficult to replicate results with a different code, since that requires good working knowledge of the original code to understand the details of the simulation and expertise in the second code to set up a comparable simulation.\cite{Schappals:2017:4270}.

These barriers impede all aspects of the \ac{CMS} fields. The learning curve to become proficient with \ac{CMS} limits the use of the tools and makes it more difficult to use multiple codes to address larger and more complex problems. Leveraging previous work to accelerate current work is less likely when the previous results cannot be reproduced. Both the learning curve and the difficulty with reproducibility make it more difficult for researchers to use the best tool for the problem. It is hard to compare codes, and switching to a new code is costly, so there is a tendency to continue using known codes even if they not as effective for a particular problem or if they are computationally more expensive. These barriers also slow the development of new and better codes for similar reasons.

Despite the issues, the use of \ac{CMS} codes is increasing with time. Any steps that reduce the learning curve, let users access more of the wide range of powerful tools available, install the software more easily, and make results more reproducible and easier to compare between codes will benefit users and further accelerate the use of these codes on significant scientific and engineering problems. This will also benefit the community developing the codes, challenging them to strive for more capable and more efficient software.

%--------------------------------------
\subsection{Existing Solutions}
%--------------------------------------
There are many efforts and codes that address the issues outlined above, indeed more than can be covered in detail here. This section will give examples of the most commonly used codes and how they fit the practical outline of how computational campaigns are structured. Starting at the lowest level, that is, the simulation codes themselves, progress is being made in standardizing parts of the input for quantum chemistry codes. For example, \texttt{libxc}\cite{libxc} is a library of \ac{DFT} functionals currently used by more than 40 \ac{CMS} software packages. \texttt{libxc} provides standard names for more than 600 functionals and ensures that the implementation is identical. Similarly, the \ac{BSE}\cite{bse} is an open library of Gaussian basis sets for quantum chemistry. The \ac{BSE} defines a standard naming and implementation for more than 700 basis sets. \texttt{libxc} and the \ac{BSE} are steps in the right direction but account for only two of the input parameters to quantum codes, although both parameters are key and have many possible values. The KIM project\cite{tadmor:2011:17} and the associated OpenKIM website\cite{OpenKIM:2025} are examples of efforts to standardize the atomistic potentials used by \ac{MD} codes for materials science.

There are a number of codes for constructing and visualizing molecular and crystal structures. Some examples are JMol/JSMol\cite{jmol}, PyMol\cite{pymol}, RASMOL\cite{sayle1995rasmol}, VMD\cite{vmd}, OVITO\cite{ovito}, VESTA\cite{vesta},  and CrystalMaker\cite{crystalmaker}. There is considerable variety in these codes. Some focus on molecular structures; others focus on crystal structures; and some on visualizing large-scale \ac{MD} runs. Most are open-source, but CrystalMaker and PyMol are commercial products. Most, but not all, of PyMol is released under an open-source license. Some have tools for building structures, and a few can create the input files to run simulations with select codes. Building and visualizing structures is a key part of most simulations.

Another group of tools assist with setting up the input for calculations for a specific code or group of codes. Most can visualize systems, have building and editing tools for the structures, and provide a \ac{GUI} for setting the parameters for the calculation. Avogadro\cite{hanwell2012avogadro,avogadro} is powerful and general, handling molecular and periodic systems, with good capabilities for building structures, analyzing results, and interfacing to a number of codes via a plug-in architecture. Other graphical interfaces support a single code, such as GaussView\cite{gaussview} for \gaussian, or for a set of similar codes, for example, Gabedit\cite{gabedit1,gabedit2}. These codes are designed to handle one calculation at a time. A structure is read from a file or created using a builder, and then the calculation is set up and run. Many of the codes can read the output of the calculation and display the results graphically. For example, many of these tools for quantum codes can display orbitals and electron densities. The combination of graphical builders with GUIs to run calculations is an excellent approach for exploration and getting started in a new project. An advantage of setting up a calculation for a specific structure is that the interface can be tailored to the problem at hand, offering only relevant choices. However, using this type of code is tedious and error-prone for running calculations on more than a few structures, since the calculation must be manually set up and run for each new system. These codes handle more of the tasks needed for a computational campaign, but only cover some of the needed tasks.

A final class of codes, or environments composed of several codes, handle running many identical or similar calculations. Most of these codes provide a library of functions for common modeling tasks such as reading or building structures and running simulation codes. The user writes e.g. a Python script that sets the parameters and calls the library functions to run the calculations. This approach avoids the labor of repetitively setting up calculations for each structure and reduces the errors in manual approaches. Examples of these codes are AFLOW\cite{Curtarolo:2012:218}, AiiDA\cite{aiida1,aiida2}, \ac{AQME}\cite{aqme}, \ac{ASE}\cite{Larsen:2017:273002}, atomate\cite{Mathew:2017:140}, BigChem\cite{Hicks:2024:142501}, Digichem\cite{digichem}, \mosdef\cite{Thompson:2020:e1742938}, pyiron\cite{Janssen:2019:24}, \qca\cite{Smith:2021:e1491}, QMFlows\cite{Zapata:2019:3191}, and wfl.\cite{Gelvzinyte:2023:124801} Some of these codes, such as AiiDA and \qca\ store results in a database, organizing the results for easy access. Databases are a very powerful tool for organizing extensive datasets, but are a challenge for most scientists to install and maintain, so the database-centric environments are most useful for expert groups preparing large datasets for machine learning, forcefield development, or public databases\cite{MaterialsProject:2025,AFLOW:2025}. The environments that do not depend on a database, such as \iac{ASE}, are easier for scientists to install and use, but they leave the organization of the simulations and results to the user.
%======================================
\section{\ac{SEAMM}}\label{SEC:SEAMM}
%======================================
The goal of the \acf{SEAMM}\cite{seamm_github,seamm_documentation} as an open source software is to facilitate and enhance computational campaigns by automating many aspects such as the individual simulations, organizing the results so that they can be viewed and analyzed by any team member, ensuring reproducibility, and tracking citations for the codes and data used.
In particular, \ac{SEAMM} has the following goals:
\begin{enumerate}
    \item Improve the productivity of users of atomistic simulation codes.
    \item Make codes easier to use and less prone to errors.
    \item Expand the range of codes that researchers can use by lowering the entry barrier. 
    \item Organize all inputs and outputs of simulations both to aid users and collaborators in viewing and analyzing the results and to ensure reproducibility.
    \item Track all workflows to ensure that results across a computational campaign are consistent and comparable and can be extended as needed.
    \item Track all codes and auxiliary data such as basis sets and forcefields, used in the workflow, to help correctly cite and credit their authors.
\end{enumerate}

%======================================
\section{Example 1}\label{SEC:EXAMPLE1}
\begin{figure}[!tbph]
	\centering
	\includegraphics[scale=1]{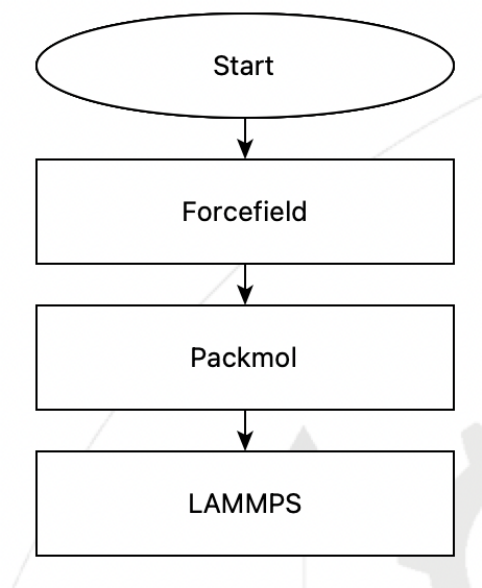}
	\caption{A sample flowchart for predicting the density of liquid ethanol via NPT MD simulation}
	\label{fig:Example 1 Flowchart}
\end{figure}

%======================================
Before covering the components and architecture of \ac{SEAMM}, a simple example will be illustrative from the user's perspective. This example will show how to use \ac{MD} in \ac{SEAMM} to predict the density of ethanol at room temperature and pressure.  At the heart of \ac{SEAMM} are flowcharts, which encapsulate a reproducible workflow as a number of steps. Figure \ref{fig:Example 1 Flowchart} shows a flowchart for creating a box of ethanol molecules as the model of the liquid and then using \ac{MD} to predict its density.

This simple flowchart has three steps: selecting the forcefield to use, building the liquid ethanol model using PACKMOL to create a cubic cell containing a few hundred molecules of ethanol, and then running a constant pressure \ac{MD} simulation using \ac{LAMMPS}.

\begin{figure}[!tbph]
	\centering
	\includegraphics[width=\textwidth]{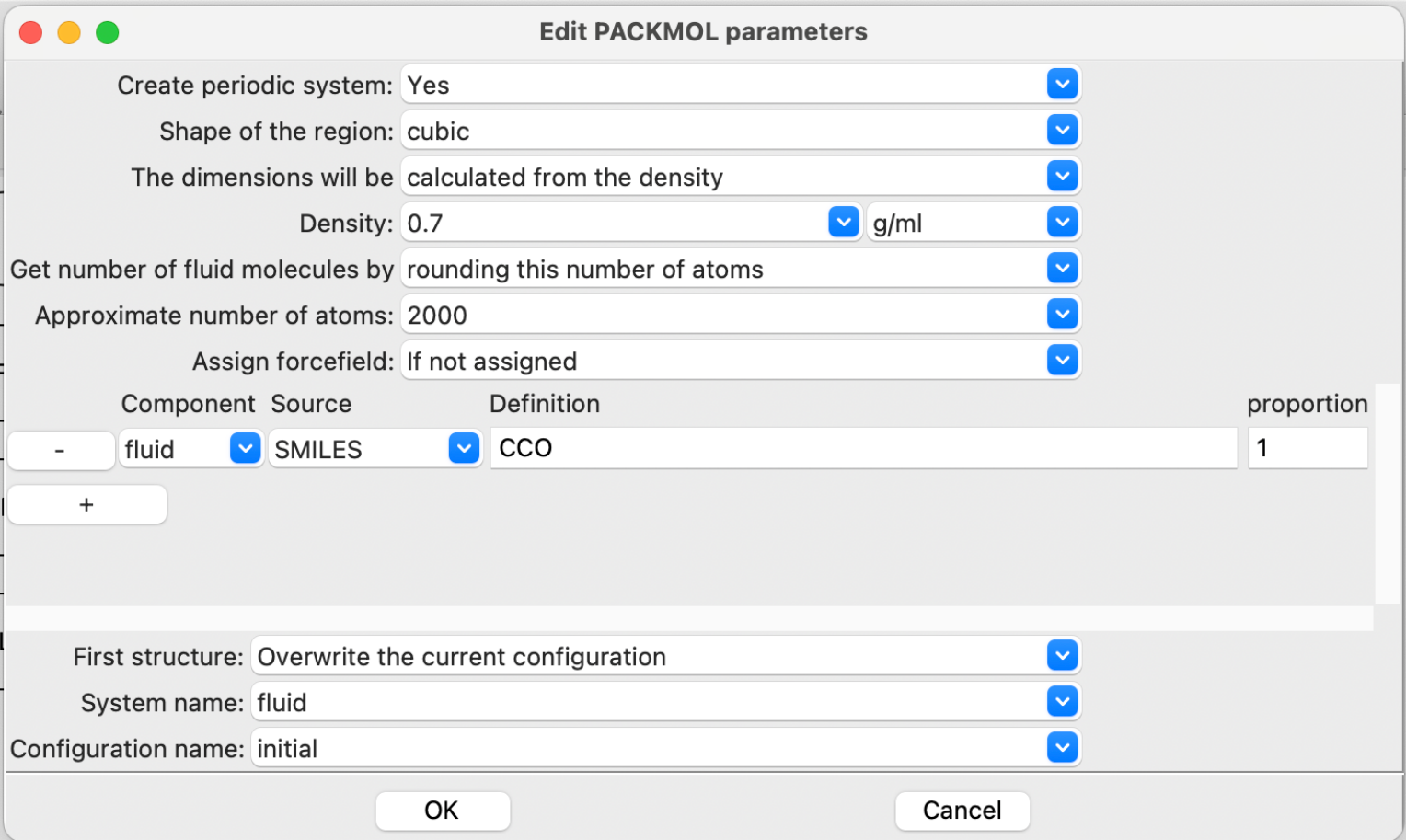}
	\caption{The PACKMOL parameter configuration dialog window in SEAMM}
	\label{fig:PACKMOL step}
\end{figure}

The user interacts with two parts of \ac{SEAMM}:
\begin{itemize}
  \item A graphical editor which can retrieve flowcharts from libraries, create them from scratch, edit them, and submit them as a job to be executed.
  \item One or more web-based dashboards that handle the execution of the flowcharts and the data storage, tracking the jobs and monitoring of the results.
\end{itemize}
The flowchart in Figure \ref{fig:Example 1 Flowchart} is created using the SEAMM \ac{GUI}. The \ac{GUI} provides a menu of steps. When the user adds a step, it is automatically added to the end of the flowchart queue, though it can be moved elsewhere if needed. The parameters for each step are edited using a dialog window provided by the step. The dialog window for configuring the PACKMOL step parameters is shown in Figure \ref{fig:PACKMOL step}.
In this example, the model is a cubic periodic cell containing a reasonable number of molecules of ethanol. The default values provided in the dialog window are very important for usability. By default, the PACKMOL step builds a spherical droplet with the size calculated from the initial guess of the density specified in the dialog and the number of molecules that contain about 2000 atoms. For this simulation, the first parameter in the dialog window should be changed for a cubic periodic cell, but the default number of molecules is already set to a reasonable value for the selected cell type. It is well-known that many properties from fluid simulations are sensitive to the cell size, converging as the size of the cell tends to infinity. The PACKMOL step uses this knowledge to guide and help users adopt good practices in their simulations. If the task was to compare the density of a number of compounds, say linear alcohols from methanol to decanol, \ac{SEAMM} uses the number of atoms to determine the number of molecules to create cells of similar sizes. On the other hand, if different simulations used the same number of molecules, larger molecules would be simulated in larger cells which introduce systematic errors in the resulting comparisons. The design of the dialog window and the default values in it guide users towards simulations that will provide reasonable results with minimum computational costs.
\begin{figure}[!htbp]
    \begin{minted}
    [
    frame=lines,
    framesep=2mm,
    baselinestretch=1.2,
    fontsize=\footnotesize,
    linenos,
    numbersep=-10pt
    ]{python}
    tolerance 2.0
    output packmol.pdb
    filetype pdb
    connect yes
    structure input_1.pdb
        inside cube 1.0000 1.0000 1.0000 26.9492
        number 222
    end structure
    \end{minted}
    \caption{A sample input for PACKMOL}
    \label{FIG:PACKMOL}
\end{figure}
Although the PACKMOL step in the \ac{SEAMM} flowchart creates the input (Figure \ref{FIG:PACKMOL}) for running PACKMOL for the user, its interpretation requires a detailed knowledge of PACKMOL. Furthermore, the specific types of molecular structure files for input and output (here, PDB files) should be provided to PACKMOL by the user. In this example, \ac{SEAMM} reads the \ac{SMILES}\cite{Weininger:1988:31} of ethanol, \ce{CCO}, generates the system model for subsequent steps. The PACKMOL step in \ac{SEAMM} also prevents another issue. PACKMOL does not create periodic cells. The documentation for PACKMOL recommends creating a somewhat smaller cell of the desired shape and transforming the created structure into a periodic system of the correct size. Using a somewhat smaller structure avoids creating atoms that are too close to each other when the system is transformed into a periodic cell. For ethanol, the PACKMOL step in \ac{SEAMM} specifies a cube of dimension 26.9492 \AA, which is then transformed into a cubic periodic cell of dimension 28.9492 \AA\ containing 222 ethanol molecules. The generated model system has the initial guess for the density of 0.7 g mL$^{-1}$ based on the user's request.

\begin{figure}[!tbph]
	\centering
	\includegraphics[width=\textwidth]{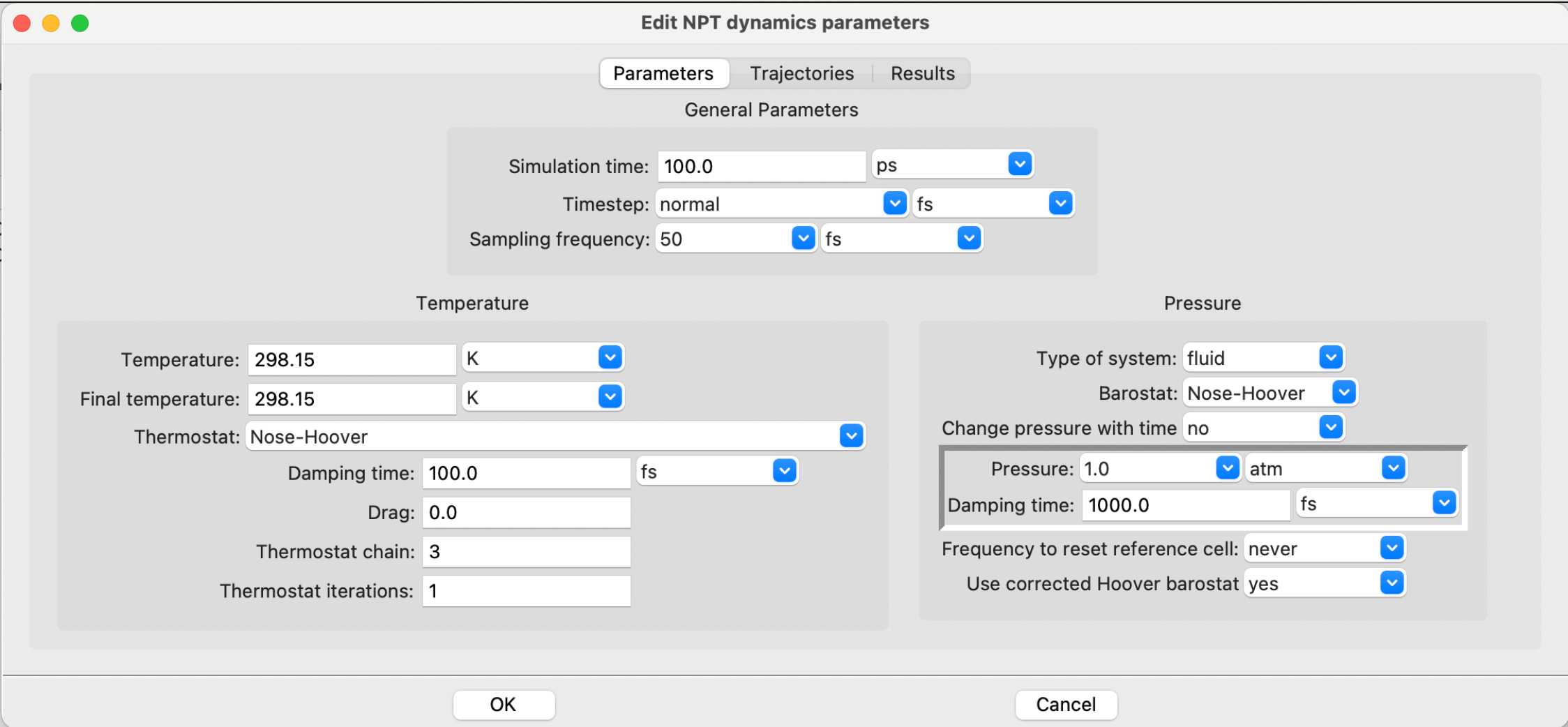}
	\caption{The parameters for \ac{LAMMPS}.}
	\label{fig:LAMMPS step}
\end{figure}

Controlling the \ac{LAMMPS} step is similarly straightforward, with a dialog window for setting up the NPT dynamics as shown in Figure \ref{fig:LAMMPS step}.
Similarly to the PACKMOL step, the \ac{LAMMPS} step also conceals considerable complexity and guides the user to a reasonable simulation. The input for \ac{LAMMPS} is not shown here because the simulation control section in it is 115 lines of text long and the data describing the molecular structure and the associated forcefield parameters add 9415 lines to it. All input and output files from the \ac{LAMMPS} step are available in the \si.

Once the flowchart is ready, the user can submit it as a job to any accessible running \ac{SEAMM} Dashboard. The Dashboard may be running on the local machine or a remote powerful high-performance computing cluster intended for performing large-scale simulations. The job can be monitored on the Dashboard as it runs and the results examined as they become available. For analyzing the job results, \ac{SEAMM} offers two useful view modes: a textual summary output, shown in Figure \ref{FIG:NPT}, and the diagrams of state variables and calculated properties, such as line plots of the density and autocorrelation function, shown in Figure \ref{fig:density plot}.

\begin{figure}[!htbp]
    \begin{minted}
    [
    frame=lines,
    framesep=2mm,
    baselinestretch=1.2,
    fontsize=\scriptsize,
    linenos,
    numbersep=-10pt
    ]{python}
Step 3: LAMMPS  2025.4.9
   LAMMPS using MPI with 4 processes.

    Step 3.2: Velocities
        Set the velocities to give a temperature 298.15 K by using a random
        distribution. LAMMPS will remove translational but not rotational momentum
        The random number generator will be initialized with the seed '1692940359'.

    Step 3.3: NPT dynamics
        100.0 ps of canonical (NPT) dynamics at 298.15 K using a timestep of 1.0 fs
        using the oplsaa+ forcefield. The temperature will be controlled using a
        Nose-Hoover thermostat. The thermostat will use a chain of 3 thermostats
        with 1 subcycles and a drag factor of 0.0. The pressure will be controlled
        using a Nose-Hoover barostat.

 
       The run will be 100,000 steps of dynamics sampled every 50 steps.
 
       Analysis of npt_state.trj
                                             Properties                                      
         Property       Value         StdErr  Units       convergence       tau    inefficiency
        ----------  ---------  ---  --------  --------  -------------  --------  --------------
            T         298.113   ±      0.215  K               1.00 ps   53.1 fs             3.1
            P          13.032   ±     29.292  atm             0.00 fs   25.0 fs             1.0
            V*      21464.385   ±     26.524  Å^3            31.00 ps  415.3 fs            17.6
         density*       0.791   ±      0.001  g/mL           28.00 ps  430.8 fs            18.2
            a*         27.793   ±      0.013  Å              28.00 ps  431.5 fs            18.3
            b*         27.793   ±      0.013  Å              28.00 ps  431.5 fs            18.3
            c*         27.793   ±      0.013  Å              28.00 ps  431.5 fs            18.3
           Etot      2017.643   ±      3.312  kcal/mol       23.00 ps  124.3 fs             6.0
           Eke       1774.646   ±      1.276  kcal/mol        1.00 ps   53.1 fs             3.1
           Epe*       243.060   ±      4.770  kcal/mol       37.00 ps  369.4 fs            15.8
          Epair*    -1172.519   ±      3.049  kcal/mol       81.00 ps  201.4 fs             9.1

          * this property has less than 100 independent samples, so may not be accurate.
          
        The structure was added as a new configuration of the current system
        named '' / 'simulated with oplsaa+'.
    \end{minted}
    \caption{The summary output from the \ac{LAMMPS} step.}
    \label{FIG:NPT}
\end{figure}
The summary output view is available for any step in a flowchart. It starts with a brief description of the main operation performed within the step alongside the control parameters that define the simulation. Next, a short summary of key results, are presented. In the example above, a table of the properties, derived from the NPT simulation along with the statistical error bars, an estimate of when the property converged to a steady state, the time constant of the \ac{ACF}, and statistical inefficiency of the sampling, are provided in the results section of the summary output. The calculated density value of 0.791 $\pm$ 0.001 kg L$^{-1}$ at 298 K and 1 atm is within about 1\% of the experimental value of 0.79,\cite{pubchem-ethanol} using a high-quality forcefield such as \ac{OPLS-AA}.\cite{Jorgensen:2005:6665}

In addition to the summary outputs, the Dashboard often offers other files that can provide more graphical details on property analysis or structure visualization. In this example, \ac{SEAMM} creates a diagram for each calculated property. The left panel in Figure \ref{fig:density plot} shows the line plot of the density simulation versus the time steps. Note the solid black line in the density diagram, indicating the average density calculated over the steady state simulation interval between 28-100 ps. The average density value is also presented in the textual output summary shown in Figure \ref{FIG:NPT} above. The right panel in Figure \ref{fig:density plot} shows the \ac{ACF} of the density in red with the error bounds shaded. The dark gray line is the best exponential fit to the \ac{ACF} corresponding to the simulation time period reported in the summary output. These diagrams provide a rapid qualitative check on the calculation results, presented in the summary output. Furthermore, the \ac{ACF} and the accompanying statistical inefficiency metric, which is a measure of oversampling of the trajectory, provide key information to improve the simulation, if necessary.
\begin{figure}[!tbph]
	\centering
	\includegraphics[width=0.8\textwidth]{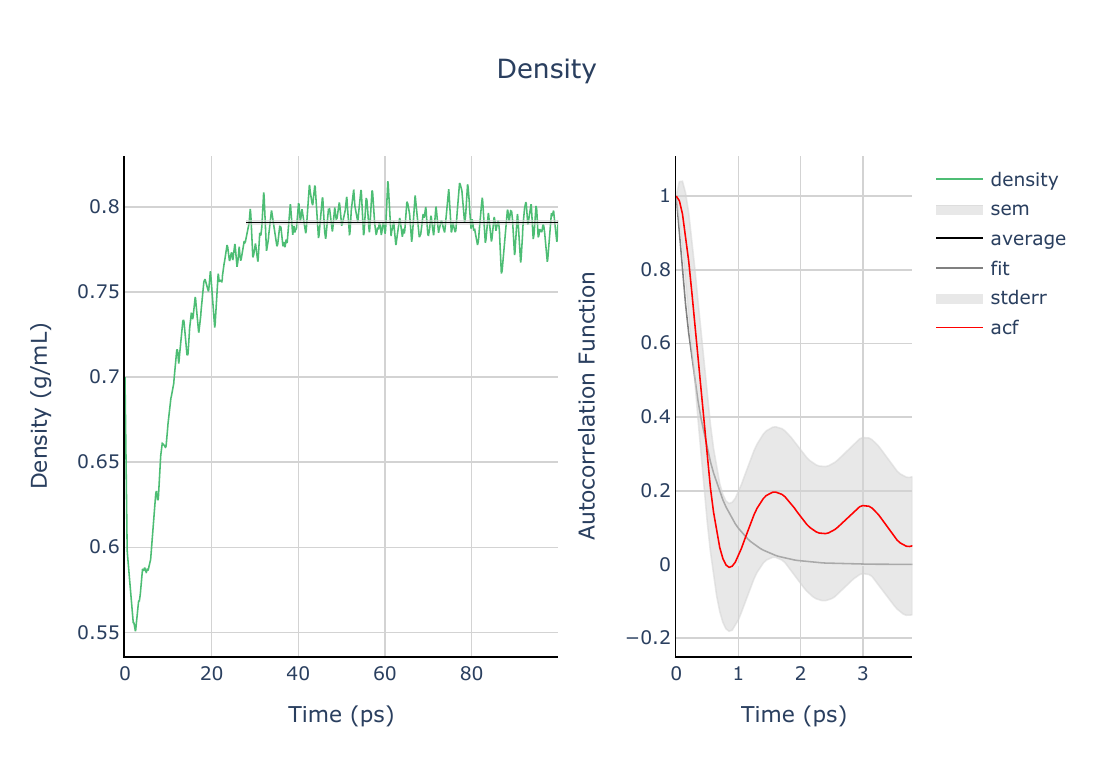}
	\caption{The line plots of the density (left) and the autocorrelation function (right) versus time, generated by the \ac{LAMMPS} step in \ac{SEAMM}. Note the horizontal line which \ac{SEAMM} automatically creates on the density plot to represent the mean density value on the plateu.}
	\label{fig:density plot}
\end{figure}

Note that by changing the \ac{SMILES} string in the PACKMOL step, the flowchart can be applied to other liquids, and mixtures can be simulated by adding additional molecules to the component table. \ac{SEAMM} supports runtime variables in flowcharts, which are set on-the-fly when submitting the job and can be used anywhere in the flowchart instead of pre-set constants. Generalizing flowcharts using runtime variables enables \ac{SEAMM} to predict the density of any liquid consisting of any molecule that the forcefield can handle.

\ac{SEAMM} and the plug-ins provide the summaries of the steps in the workflow, as well as appropriate citations for the codes and parameter sets used, both as text and as \bibtex. This feature reduces the required effort for documenting the simulation details and providing a comprehensive description of the workflow. In the example shown above, the output and citations can easily be converted into the following sample description for scholarly publications: \textit{The \ac{SEAMM} environment\cite{seamm,lammps_step,packmol_step} was used to predict the density of ethanol at 298.15 K and 1 atm via an \ac{MD} simulation with an isothermal-isobaric (NPT) ensemble using a Nose-Hoover thermostat/barostat\cite{Shinoda:2004:134103}, running for 100 ps with a 1 fs time step using time-reversible, measure-preserving Verlet and rRESPA integrators\cite{Tuckerman:2006:5629} with the \ac{LAMMPS} code.\cite{plimpton:1995:1,lammps1} The potential energy of the system was calculated using the \ac{OPLS-AA} forcefield\cite{Jorgensen:2005:6665} with custom parameters for ethanol obtained from the LigParGen website\cite{Dodda:2017:W331,LFQSCWFLJHTTHZ-UHFFFAOYSA-N,Dodda:2017:3864}. Non-bonded interactions were cut off at 10 \AA\ with long-range Coulomb terms handled using a particle-particle particle-mesh solver\cite{hockney1988computer} and a tail correction\cite{sun:1998:7338} for long-range Lennard-Jones terms. The model of liquid ethanol was a cubic periodic cell containing 222 molecules of ethanol. The initial structure was created using PACKMOL \cite{Martinez:2009:2157,packmol} to create a cubic cell with a density of 0.7 g/cm$^3$ using an ethanol structure created from \ac{SMILES} using OpenBabel\cite{OBoyle:2011:33, obabel}. The analysis of the trajectory was carried out by the LAMMPS step\cite{lammps_step} in \ac{SEAMM}, using the pymbar library\cite{pymbar,Chodera:2016:1799,Shirts:2008:124105,Chodera:2007:26} to determine when the system had reached a steady state and to perform the statistical analysis of that portion of the trajectory.} 

%======================================
\section{\ac{SEAMM} Components}\label{SEC:COMPONENTS}
\begin{figure}[H]
	\centering
	\includegraphics[width=\textwidth]{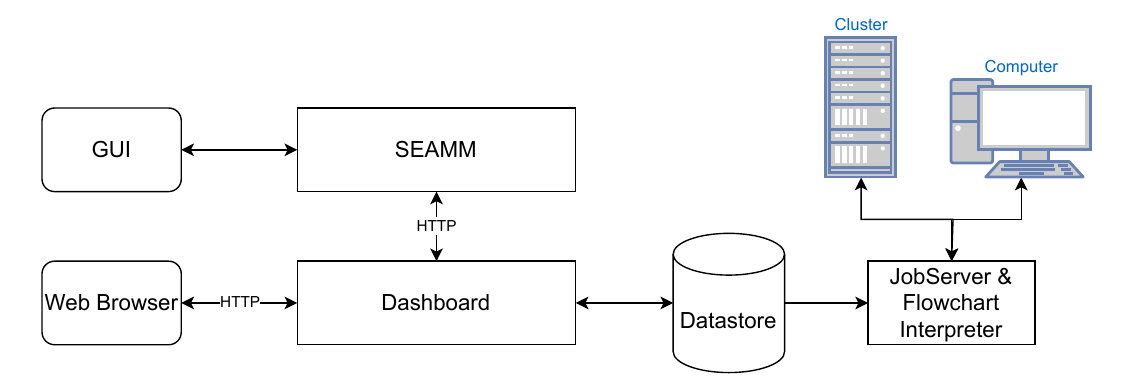}
	\caption{Components of the \ac{SEAMM} environment.}
	\label{fig:SEAMM Architecture}
\end{figure}

%======================================
Figure \ref{fig:SEAMM Architecture} provides a high-level view of various parts of the \ac{SEAMM} environment. The next few subsections will provide more details on each components and how individual parts work together to provide a comprehensive software ecosystem for performing \ac{CMS} simulations.

%--------------------------------------
\subsection{SEAMM Core}\label{SUBSEC:SEAMMCORE}
\begin{figure}[!tbph]
	\centering
	\includegraphics[scale=0.9]{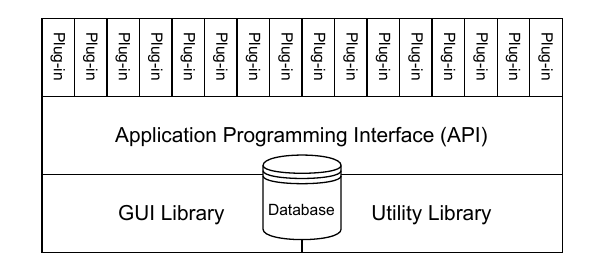}
	\caption{High-level software stack view of \ac{SEAMM} core component}
	\label{fig:SEAMM Framework}
\end{figure}

%--------------------------------------
Figure \ref{fig:SEAMM Framework} shows \ac{SEAMM}, which is the central part of the \ac{SEAMM} environment in Figure \ref{fig:SEAMM Architecture}, in more detail. There are five parts: a database for storing computational results during a job, a \ac{GUI} and utility libraries that contain useful functionality for developers, an \ac{API} that defines the interface between \ac{SEAMM} and the plug-ins, which provide all the functionality that users see, e.g. the flowchart and dialogs in Example 1.

%--------------------------------------
\subsection{Plug-ins}\label{SUBSEC:PLUGINS}
%--------------------------------------
\ac{SEAMM}, at its core, is a framework that provides its own data model and numerous utility functions without offering any end-user functionality. In order to enhance modularity, \ac{SEAMM} directs external end-user functionalities into separate code units, called plug-ins, as  shown in Figure \ref{fig:SEAMM Framework}. Each plug-in corresponds to a step in a flowchart. Plug-ins provide interfaces to external \ac{CMS} simulation tools, builders for creating systems, analysis modules, control steps in flowcharts such as loops and sub-flowcharts, and other functionality that users need to run their simulations. Plug-ins are developed based on \ac{SEAMM} \ac{API} but are independent software modules and distributed separately from \ac{SEAMM}. \ac{SEAMM} installer finds the plug-ins, published on \ac{PyPI}\cite{pypi}, and installs or updates them upon user request.

In Example 1, the three steps in the flowchart in Figure \ref{fig:Example 1 Flowchart} correspond to individual plug-ins that provide interfaces for the PACKMOL, \ac{LAMMPS} and the forcefield steps. When the flowchart is executed, the forcefield step plug-in reads the forcefield from a disk file and stores it in internal data structures for easy access. A substantial part of the utility library is devoted to handling forcefields, including automatic ``atom typing'' and preparing the lists of forcefield parameters for codes such as \ac{LAMMPS}. As such, the forcefield plug-in can automate most aspects of forcefield development for both users and developers, as detailed in Example 1, where the user only had to choose which forcefield to they want to use and everything else related to the forcefield was hidden.

%--------------------------------------
\subsection{Data Model and Internal Database}
\label{SUBSEC:DATAMODEL}
%--------------------------------------
\ac{SEAMM} supports simulations of molecules, fluids, and amorphous and crystalline solids plus interfaces between different phases. Although these structures are closely related, different disciplines use different terminologies which can be confusing. Before proceeding with discussing data storage, we will cover the adopted terminology in \ac{SEAMM}. The term ``\textit{system}'' denotes a collection of atoms, molecules, and/or crystals that are simulated, analogous to experiments on thermodynamic systems.\cite{thermodynamic_system} In \ac{SEAMM}, the \textit{system} itself is an abstraction that consists of one or more \textit{configurations}, which themselves are composed of atoms along with their coordinates, optional bonds between atoms, the periodic cell for periodic \textit{systems}, the point- or space-group symmetries, and the charge and spin/magnetic states of the \textit{system}. Many codes call what SEAMM refers to as a \textit{configuration} a ``molecule'' or ``structure'', or in some cases they are ``conformers'' or ``frames'' in a \ac{MD} trajectory. The \textit{configurations} in \ac{SEAMM} are more general than the structures or the frames in trajectories in most codes. Different \textit{configurations} of a \textit{system} can have different atoms and different bonds. With this flexibility, SEAMM can support reactive dynamics and other simulations of chemical reactions in a natural way. \ac{SEAMM} can also support grand canonical ensembles, where the number of atoms or molecules can change such as when simulating vapor deposition by creating a stream of new molecules to impinge on a target surface.

Most quantum codes take the charge and spin as input parameters for the simulation. In \ac{SEAMM}, charge and spin multiplicity are properties of the \textit{configurations}. This feature extends the usability and generality of flowcharts to handle neutral molecules, anions, cations, and different spin states within the same framework. Chemists think of singlet and triplet oxygen as two different chemical entities, as does \ac{SEAMM}.

\ac{SEAMM} uses the concept of a \textit{subset}, which is an arbitrary collection of atoms in a \textit{configuration}. Atoms can be in more than one \textit{subset}, which allows \textit{subsets} to be defined for commonly used concepts such as residues and chains in proteins or adsorbents and adsorbates. \textit{Templates} are closely related to \textit{subsets} but contain a \textit{configuration} that has atoms, bonds, and coordinates, as well as other useful attributes such as names, partial charges, and forcefield atom types. A \textit{template} is a prototype that can be used to find identical substructures in a \textit{system}, matching them atom-by-atom regardless of the order of the atoms. Once matched, attributes such as atom names, partial charges,  and bond orders can be transferred from the \textit{template} to the substructure. For example, \textit{templates} of the amino acids could be used to match the different residues in a protein and provide the structure with standard names, partial charges, and atom types.

The data for \textit{systems}, \textit{configurations}, \textit{templates}, calculated properties and other results are stored in an internal relational database. Rather than accessing the database directly, the utility library provides a facade over the database, so to the rest of the \ac{SEAMM} and to the plug-ins, the underlying data model is accessed through an object-oriented framework with \textit{systems}, \textit{configurations}, atoms, bonds, etc. This model is common in simulation codes, so developers will probably find this aspect of the \ac{SEAMM} similar to other software packages they may already be familiar with.

There are several advantages to using a relational database. \acp{RDBMS} constitute a mature, well-understood, high-performance and scalable technology. \ac{SEAMM} currently uses SQLite\cite{sqlite} for database management because it is small and lightweight, well-integrated with Python, and designed for use within programs. Since \ac{SQL} is a standard language for \acp{RDBMS}, including SQLite,  \ac{SEAMM} could switch to a different \ac{SQL}-based \ac{RDBMS} if needed. A second advantage, particularly of SQLite, is that the database file is portable across platforms, storing data with full precision fidelity. As such, \ac{SEAMM} and the database files are also portable across different hardware and operating systems. Furthermore, the most important advantage gained is the flexibility and extensibility that a database provides. 

The schema used in the database defines the data that it stores as tables, columns, and relations between them. There are two main parts to the schema used by \ac{SEAMM}: a conventional schema for \textit{system} and \textit{configurations}, and a star schema\cite{star_schema} for property data and simulation results.

\begin{sidewaysfigure}
	\centering
	\includegraphics[width=.95\textheight,height=.63\textwidth]{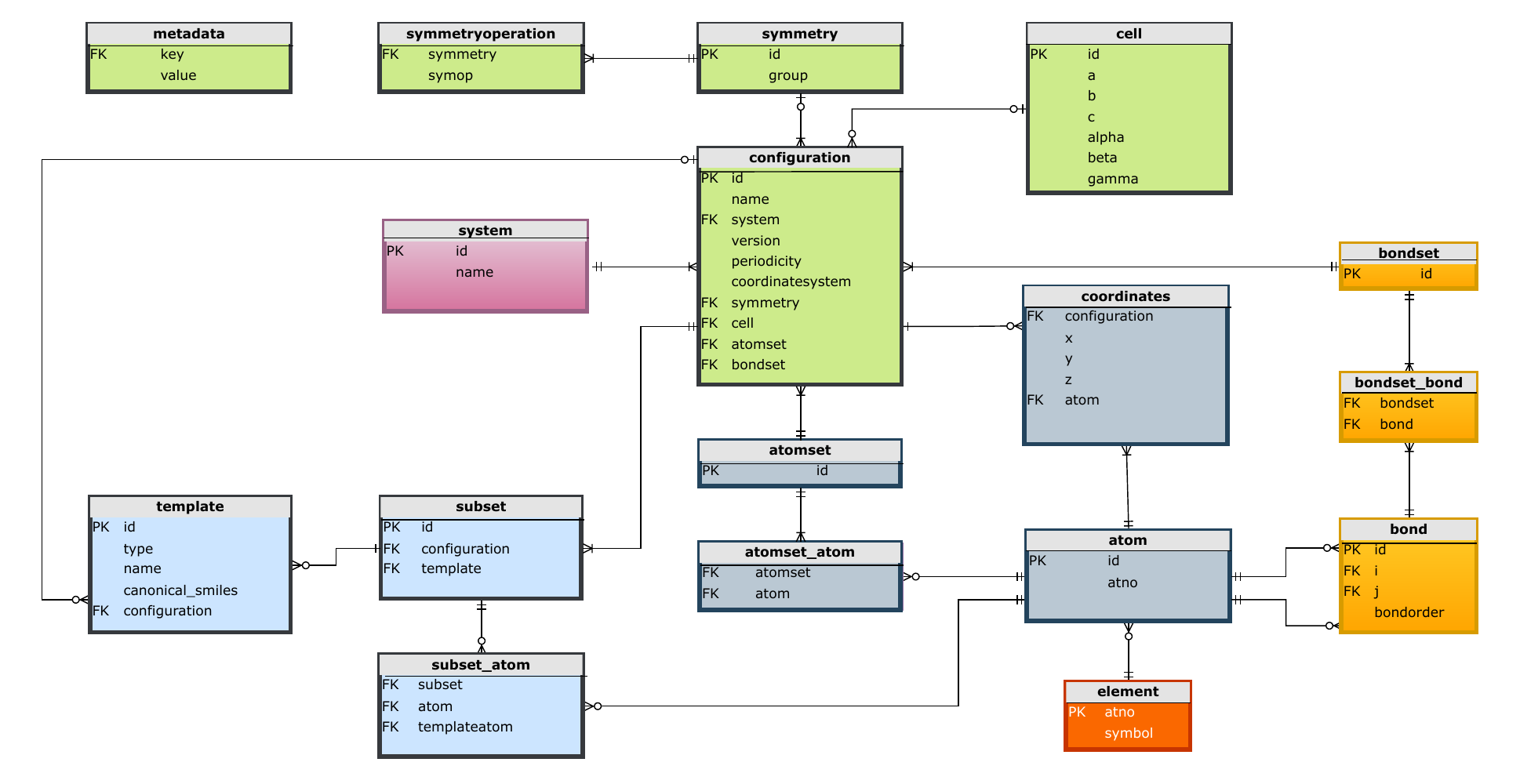}
	\caption{Database schema for molecular structures}
	\label{fig:SystemDB schema}
\end{sidewaysfigure}

Figure \ref{fig:SystemDB schema} shows the schema for \textit{system} and \textit{configurations}. The \textit{configuration} is the heart of the description, connected to the atoms and bonds through the junction tables \texttt{atomset\_atom} and \texttt{bondset\_bond}. This allows \textit{configurations} to contain different atoms or bonds, while avoid duplicating the shared atoms and bonds for storage size reduction. The data storage size is important for large \ac{MD} simulations with extensive trajectories. Storing the atomic coordinates separately from other atomic features also helps limit the amount of data stored during \ac{MD} simulations, since in many simulations, the topology of the structure-- atom types and bonds-- do not change, but the atomic positions change with every step. The remainder of the schema is the handling of the point or spacegroup symmetry and periodic cells with three tables at the top, and \textit{subsets} and \textit{templates} in the lower left portion of the diagram.

\begin{figure}[!tbph]
	\centering
	\includegraphics[width=\textwidth]{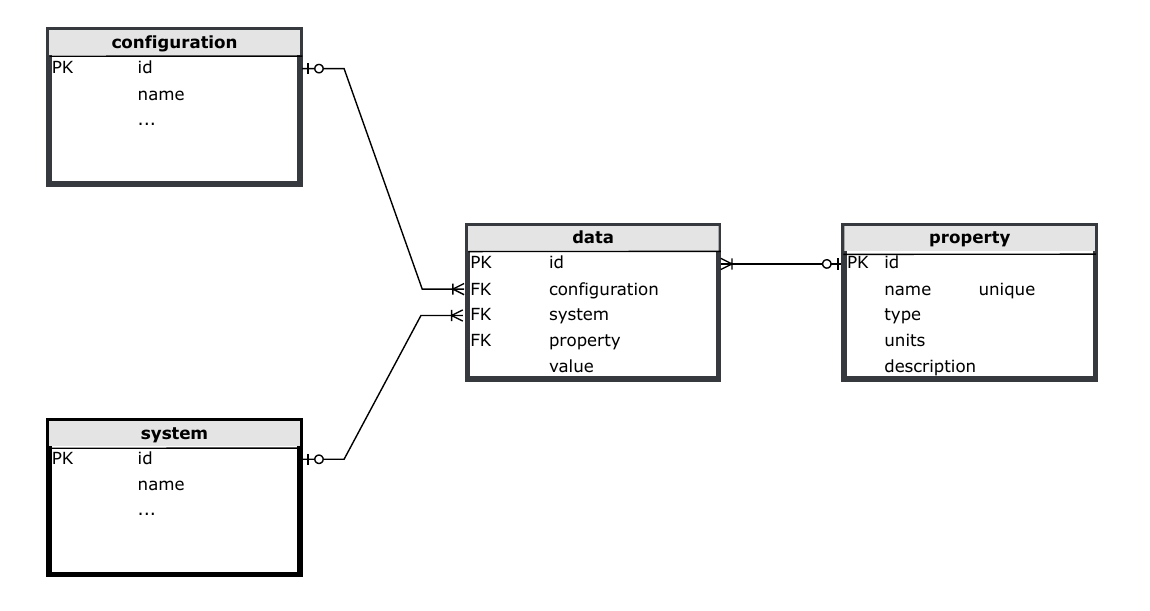}
	\caption{Star schema for system properties}
	\label{fig:star schema}
\end{figure} 

The second part of the schema is the star schema for handling properties and calculated results, shown in Figure \ref{fig:star schema}. The values of the properties and other results are stored in the \texttt{data} table. In a conventional schema, the table would have a column for each property, such as energy, dipole moment, band gap, etc. This is not practical because the number of different results from all possible simulations is large but each specific simulation generates only a small number of different results. This would require an extremely wide table that would only be sparesly populated. Adding a new type of result would require changing the schema, versions of database, \etc\ which is unsustainable. 

The star schema circumvents this problem by using metadata to describe the data in the \texttt{data} table and storing the metadata in the \texttt{property} table. As such, the \texttt{data} table remains dense. Adding a new property or type is also possbile by insterting a new row of metadata to the \texttt{property} table, which can be done while \ac{SEAMM} is running, without the need for any change to the code requiring a new release. The disadvantage of the star schema is that finding data requires several layers of indirection or ``join'' operations in \ac{SQL} vocabulary. Fortunately, \acp{RDBMS} are highly optimized for performing join operations because star schemas are widely used in commerce to store data, such as credit card transactions or airline reservations, which require high-performance and truly massive databases but similarly sparse data.

Similar to the database that stores the structures, \ac{SEAMM} provides an object-oriented facade over the star schema. The facade provides methods to define new properties, to add property data to \textit{systems} and \textit{configurations}, and to search for and retrieve the data. \ac{SEAMM} defines standard properties such as electronic energy, formation enthalpy, dipole moment, volume or band gap, providing a \defacto\ standard. The plug-ins can dynamically add any other properties that they need, such as thermal expansion coefficient, heat capacity or viscosity, though if another plug-in needs the data, the plug-ins must agree on a consistent and common naming convention.

It is impossible to overstate the importance of flexibility and extensibility of the data management systems for handling a wide range of tools and applications in the \ac{CMS} fields. It is also key in ensuring \ac{SEAMM}'s ability to evolve, support new codes and enable new science for many decades to come.

%--------------------------------------
\subsection{Utility Library}\label{SUBSEC:UTILITY}
%--------------------------------------
The utility library contains a number of useful tools that can be assembled by plug-ins to meet specific needs such as:
\begin{enumerate}
    \item A comprehensive package for handling units, based on the open-source Python package Pint\cite{pint} and customized to handle unit conversions that are common in chemistry and materials science;
    \item interfaces to Open Babel\cite{OpenBabel}, RDKit,\cite{RDKit} \ac{ASE}\cite{Larsen:2017:273002}, and geomeTRIC\cite{geoMETRIC} to facilitate their use within the \ac{SEAMM} environment;
    \item functions to traverse the molecular graphs defined by the bonds between atoms to find individual molecules, neighbors of atoms, angles, dihedrals, and other topological features of molecules;
    \item handling of the forcefield files;  and
    \item a flexible printing system modeled after the Python logging system.
\end{enumerate}

%--------------------------------------
\subsection{\ac{GUI} Library}\label{SUBSEC:GUI}
%--------------------------------------
The \ac{GUI} in \ac{SEAMM} is implemented using a \ac{MVC} design pattern, which allows multiple graphical implementations to be used interchangeably. The initial implementation is based on tkinter\cite{tkinter}, which is Python's native \ac{GUI} package. The \ac{GUI} library in \ac{SEAMM} provides a number of widgets that are tailored to the needs of \ac{CMS}, such as periodic tables and entry fields that have built-in unit handling. These features make the development of the \acp{GUI} for \ac{SEAMM} plug-ins easier and help standardize the resulting user interface and experience. There is also support for creating diagrams similar to those shown in Figure \ref{fig:density plot}.

%--------------------------------------
\subsection{The \ac{SEAMM} \acl{API} for Creating Plug-ins}
\label{SUBSEC:API}
%--------------------------------------
Plug-ins consist of at least four components: a computational class that defines the core computational functionality of the plug-in; a graphical class that defines the \acf{GUI}; a class that contains the parameters which are set by the \ac{GUI} and consumed by the computational class; and a small factory class that \ac{SEAMM} uses to instantiate objects of the computational and graphical classes. The first three classes are subclassed from base classes provided by \ac{SEAMM} core. The factory class is minimal, consisting of two methods, each of which has a single line of code that creates and returns an instance of the compute or graphical class for the plug-in. The three base classes plus the factory class define and enforce much of the \ac{API} for \ac{SEAMM}. Python does not directly support virtual classes; however, base classes raise exceptions in methods that must be overridden in the subclass. These base classes also contain a number of methods that implement a significant portion of the functionality of a plug-in. In most cases, these methods are sufficient out-of-the-box, but can also be overridden for special cases.

\ac{SEAMM} provides a Python Cookiecutter\cite{seamm-cookiecutter} which is built upon the Cookiecutter for Computational Molecular Sciences.\cite{cms-cookiecutter} The \ac{SEAMM} Cookiecutter creates a directory and project for a new \ac{SEAMM} plug-in. It is populated with the initial source files for the plug-in, a skeleton of the documentation, tests, and the files needed to create a complete project in GitHub including the GitHub Actions to check the code, run tests, and eventually release on \ac{PyPI}. As such, the Cookiecutter greatly lowers the development barrier for creating plug-ins \ac{SEAMM} and ensures their consistency with the \ac{SEAMM} \ac{API}. The initial skeleton, created by the \ac{SEAMM} Cookiecutter can be installed and used immediately. However, it has no useful functionality on its own. Therefore, the developer's task is to add the required parameters, edit the dialog window if the automatic placement of the parameters is not sufficient, and create the code for executing the core function. For example, a plug-in that runs a simulation code needs to define the control parameters for it. The \ac{GUI} code will automatically display the widgets for the parameters in a table, which is sufficient if there are only a few parameters. For more complex plug-ins the developer needs to change the code that positions the widgets and impose any dependencies between parameters if the value of one parameter affects the others. The developer must also create the code to take the parameters and create the input for the simulation code, and also reformat the atomistic structure for the code. Finally, the code to run the simulation code, and read and analyze the output also needs to be written.

The developer must supply any codes, libraries, or data that are needed to run the simulation, and may need to write a simple installer for these items. \ac{SEAMM} provides templates and utilities for such installers. At the very least, the plug-in must document how users can obtain and install any required code and data. If the dependencies are open-source or accessible in the public domain, the best practice is to ensure that the plug-in retrieves and installs everything that it needs. Simulation codes should be installed in separate Conda environments or Docker containers to minimize conflicts between different codes and their dependencies.

%--------------------------------------
\subsection{Dashboard and Datastore}
\label{SUBSEC:DASHBOARD}
%--------------------------------------
Example 1 briefly showed the Dashboard from a user's point of view. The Dashboard is a custom web server for monitoring and managing jobs. The jobs are stored in the datastore, which is a combination of a \ac{RDBMS} and files on disk. The database stores information about the jobs as well as pointers to the disk files, allowing for rapid access. The inputs and outputs of the simulations are stored as files and indexed in the database. This allows users to directly access the files or use shell scripts and similar tools to access them if they wish. It also allows users to backup the files using tools that they are comfortable with, rather than trying to backup what would be a very large database if the simulation results were stored therein. The information in the database for each job is also saved in a disk file alongside other job files. The Dashboard can recreate the database from these files if it is corrupted.

The Dashboard provides a \ac{REST} \ac{API} to the datastore, allowing programmatic access to the data. The \ac{SEAMM} \ac{GUI} uses this \ac{API} when submitting jobs to the Dashboard. Since this feature uses hypertext transfer protocol (HTTP) for data transport, \ac{SEAMM} can access the Dashboard from anywhere via a browser. Users can also access the Dashboard and jobs on any machine that they can access directly or through \eg\ secure shell (SSH) tunneling.

%--------------------------------------
\subsection{JobServer}\label{SUBSEC:JOBSERVER}
%--------------------------------------
The JobServer is a daemon or long-running service is responsible for running jobs, submitted to the Dashboard, and placing them in the datastore queue. The JobServer accesses the datastore directly, periodically checking the queue for monitoring jobs that are waiting to run. The current version of the JobServer executes a job's flowchart directly on the machine or in a Docker image if \ac{SEAMM} is installed using Docker. Future versions will be able to work with queueing systems which are typically used on large computer clusters.

%--------------------------------------
\subsection{Flowcharts}\label{SUBSEC:FLOWCHARTS}
%--------------------------------------
Flowcharts are text files in \ac{JSON} format that define the steps in each workflow, their control parameters and the connections between them. Each step in a flowchart corresponds to a plug-in, represented by a graph node. The individual steps (nodes) are composed of three pieces of information defining the plug-in-- the Python module of the plug-in, the class name, and the version of the plug-in-- plus data that the plug-in defines and handles. This information allows a code, interpreting the flowchart, to instantiate each plug-in and have it read its internal information from the flowchart. The \ac{API} for plug-ins specifies an \texttt{edit} method, which \ac{SEAMM} uses to present the plug-in's dialog window to the user, and a \texttt{run} method, which handles the actions of the plug-in. \ac{SEAMM} also provides an interpreter to execute the flowcharts. Each flowchart starts with the appropriate shebang line pointing to the interpreter. Thus, if the flowchart file is executable, it will run if called directly within a command-line interface.

This design provides a great deal of flexibility. Because the version of each plug-in is recorded in the flowchart, it is possible to use either the latest version of the plug-ins, or the version specified in the flowchart. Since plug-ins are responsible for handling their own data in the flowchart, they are free to store their data in any convenient form, regardless of data management aspects in any other part of \ac{SEAMM}. Because flowcharts are text files, they can be shared by email, saved to disk, or placed in libraries. \ac{SEAMM} can publish flowcharts on Zenodo\cite{zenodo}, a free and open-source research repository developed by CERN and OpenAIRE, which assigns the flowcharts a \ac{DOI} for future reference. Flowcharts can also be searched on Zenodo and be downloaded directly into \ac{SEAMM}.

%======================================
\section{Example 2: Rearrangement of Methylisocyanide to Acetonitrile}
\label{SEC:EXAMPLE2}
%======================================
One of the strengths of \ac{SEAMM} is its ability to use a range of codes to tackle complex multi-scale problems in \ac{CMS}. This section will focus on the rearrangement of methylisocyanide to acetonitrile, which is a prototype for unimolecular reactions and has been studied extensively, both experimentally and computationally.\cite{liskow:1972:4509,liskow:1972:5178,redmon:1978:5386,Saxe:1980:3718, Nguyen:2018:2532} We will reproduce the results of one of the early computational studies\cite{Saxe:1980:3718} and then replicate the results using a wide range of computational methods in \ac{SEAMM}. The focus will be on the energy of the transition state structure and the curvature of the potential energy surface at the saddle point.

We will start by reproducing the optimized transition state structure and its harmonic vibrational frequencies as provided in Ref.~\citenum{Saxe:1980:3718} and then move to locate the transition state structure from scratch. Given the structure, it is straightforward to calculate the vibrational frequencies and the normal modes using quantum chemistry software such as \gaussian\cite{g16} or \psifour.\cite{Psi4} The results are shown in Table \ref{table:1}.

\begin{table}
\centering
\begin{tabular}{m{3cm} *{4}{r}}
\hline\hline
 & \mr{2}{*}{\textbf{Ref.\citenum{Saxe:1980:3718}}} & \mr{2}{*}{\textbf{\gaussian}} & \mr{2}{*}{\textbf{\psifour}} & \textbf{Finite} \\
  & & & & \textbf{Difference} \\
\hline
\multirow{11}{3cm}{\textbf{Frequencies ($\wn$)}} & 458$i$ & 458$i$ & 458$i$ & 459$i$ \\
 & 255  & 255  & 255  & 255  \\
 & 677  & 677  & 677  & 677  \\
 & 1063 & 1063 & 1063 & 1063 \\
 & 1083 & 1083 & 1083 & 1083 \\
 & 1457 & 1457 & 1457 & 1456 \\
 & 1590 & 1590 & 1590 & 1589 \\
 & 1599 & 1599 & 1599 & 1599 \\
 & 2189 & 2189 & 2189 & 2189 \\
 & 3272 & 3272 & 3272 & 3272 \\
 & 3388 & 3388 & 3388 & 3388 \\
 & 3411 & 3411 & 3411 & 3411 \\
\textbf{Energy ($\Eh$)} & -131.84756 & -131.84744 & -131.84738 & -131.84744 \\
\hline\hline
\end{tabular}
\caption{The vibrational frequencies ($\wn$) and energies ($\Eh$) of the transition-state structure for the rearrangement of methylisocyanide to acetonitrile calculated at the HF/DZP level of theory. The reference values are obtained from Ref.~\citenum{Saxe:1980:3718} }
\label{table:1}
\end{table}

The first column lists the reference frequencies obtained from Ref.~\citenum{Saxe:1980:3718}, where the second derivatives needed for the harmonic vibrational analysis were calculated using finite differences of gradients. The second and third columns in Table \ref{table:1} show the vibrational frequencies calculated using \ and \psifour, respectively, using analytic second derivatives. The analytic frequencies are identical to those of Ref.~\citenum{Saxe:1980:3718} within less than 1$\wn$ while the energies are in perfect agreement within 0.0002 $\Eh$ (or $\approx$0.13 \kcalmol). The different codes make different approximations to reduce the computational cost. For example, \psifour\ uses the \ac{RI} and various cutoffs by default when evaluating the two-center integrals. \gaussian, on the other hand, does not use \ac{RI}. As noted in Ref.~\citenum{Saxe:1980:3718}, the code used at the time could not handle the $d$ functions in the basis set. As such, the $d$ functions were approximated by off-center $p$ orbitals which introduced an error of $\approx$0.0001 $\Eh$. Note that such errors or approximations tend to be systematic. Thus, relative energies are more meaningful and accurate in this context. In this section, we have made no effort to tune the input parameters in \psifour\ and \gaussian\ and benefited from reasonable default values in \ac{SEAMM} to reduce the errors from the approximations made to gain performance. The the precision of the current results is also well within chemical accuracy and on-par with the precision of the original calculations performed more than 40 years ago. These results provide strong evidence for the reproducibility of the quantum chemical calculations between different codes with quite different implementations and running on very different computers which stood the test of
time over decades.

Both \gaussian\ and \psifour, as well as a number of other codes, with available interfaces in \ac{SEAMM}, can calculate second derivatives analytically. However, many codes may not offer this feature. The \texttt{Thermochemistry} plug-in in \ac{SEAMM} eliminates this issue by using finite differences or the first derivatives to calculate harmonic vibrational frequencies and properties. The last column in Table \ref{table:1} shows the results of the finite difference approach mentioned above using \gaussian to calculate the energy and first derivatives. Using the default step size of 0.01 \AA\ in the \texttt{Thermochemistry} plug-in, the results are essentially identical to those calculated analytically, with a deviation of 1 $\wn$ in the frequencies. Another advantage of using the \texttt{Thermochemistry} plug-in is that it ensures that the vibrational analysis is handled identically when using different methods and codes. As an added benefit, the \texttt{Thermochemistry} plug-in makes it easy to switch between different simulation codes, since only a small subflowchart, responsible for calculating the energy and forces, needs to be changed.

Beyond reproducing the results for the given transition state structure, the \texttt{Reaction Path} plug-in for \ac{SEAMM} finds approximate reaction paths and transition states using the \ac{NEB} method\cite{neb:1998,neb:2000,neb:2019,ci-neb:2000,idpp:2014}. Once an approximate transition state is located, the \texttt{Structure} plug-in can optimize it, in the process ensuring that there is one and only one mode with an imaginary frequency. Similar to the \texttt{Thermochemistry} plug-in, both plug-ins only need the energy and gradients of the structure, allowing them to be used with a wide range of simulation methods in \ac{SEAMM}. 

\begin{table}
\centering
\begin{tabular}{c | *{6}{r}}
\hline\hline
 & \textbf{HF/DZP}$^a$ & \textbf{HF/tight}$^b$ & \textbf{PM6-org}$^c$ & \textbf{PM7}$^d$ & \textbf{DFTB/3ob}$^e$ & \textbf{xTB/GFN2}$^f$\\

\hline
  \textbf{\ce{CH3NC}} & 0.0 & 0.0 & 0.0 & 0.0 & 0.0 & 0.0 \\
 \textbf{TS} & 43.6 & 42.5 & 60.5 & 73.1 & 60.2 & 62.8 \\
  \textbf{\ce{CH3NC}} & -19.5 & -20.0 & -29.5  & -20.7 & -10.4 & -23.5 \\
 \textbf{Freq.}$^g$ & 458.0$i$ & 463.0$i$ & 577.0$i$ & 696.0$i$ & 343.0$i$ & 412.0$i$ \\
 \hline\hline
 \end{tabular}
\caption{Comparison of the enthalpies (\kcalmol at 298K and 1 atm) of the structures of methylisocyanide, the transition state, and acetonitrile relative to methylisocyanide, plus the curvature of the imaginary mode at the transition state for a range of different models.}
\label{table:2}
  \begin{tablenotes}
    \footnotesize
    \item $^a$Hartree Fock with \gaussian\ using the ``DZP'' basis. $^b$Hartree Fock with \fhiaims\cite{blum:2009:2175, havu:2009:8367} using the ``tight'' basis\cite{zhang:2013:123033}. $^c$Hartree Fock with \mopac\cite{mopac} using the ``PM6-ORG'' parameterization\cite{stewart:2023:284}. $^d$Hartree Fock with \mopac\ using the ``PM7'' parameterization\cite{stewart:2012:1}. $^e$DFTB\cite{scc-dftb} with \dftb\cite{dftbplus} using the ``3ob'' parameterization\cite{3ob}. $^f$xTB\cite{xTB} with \dftb\ using the ``GFN2'' parameterization\cite{GFN2}. $^g$ Frequencies are in $\wn$.
  \end{tablenotes}
\end{table}

Table \ref{table:2} shows the relative enthalpies of the structures in the reaction path, as well as the imaginary frequency of the transition state, calculated with a range of codes and methods, ranging from \textit{ab initio|} Hartree Fock method with Gaussian orbitals and numerical ones, semiempirical Hartree Fock methods in \mopac, and semiempirical \ac{DFT} using \dftb. The \abinit\ Hartree Fock methods give almost identical results. The observed small differences are mainly due to differences in the quality of the analytical and numerical basis sets in various codes. The results presented in Table \ref{table:2} demonstrate that the semiempirical methods, whether based on Hartree Fock or \ac{DFT}, overestimate the reaction energy barrier height and have a larger variance for both the enthalpy of reaction and the imaginary frequency.

Similar calculations with ReaxFF forcefields\cite{strachan:2003:98301,singh:2013:104114,shan:2014:962} and machine learning potentials\cite{smith:2019:2903,devereux:2020:4192} could not locate the transition state, nor could the transition state be successfully optimized starting from the known transition state structure using these forcefields or potentials. Although details of the parameterization are difficult to track in these methods, it appears that the training data for all of the methods examined include cyanides but not isocyanides. Each of the potentials tried predicts a nonlinear C-N-C bond angle in methylisocyanide instead of the known linear structure, indicating that the extrapolation of the potentials to the isocyanide functional group is not reasonable. So, while \ac{SEAMM} is capable of locating the transition state and calculating its thermochemistry using such potentials, for this particular reaction none of the forcefields or machine learning potentials tried in this study could represent the potential energy surface in a physically reasonable way, because the reactant and transition state are too far from the training sets. More details about the calculations and results in this example are included in the \si, along with a description of the flowchart used.

%======================================
\section{Example 3: Advancing Understanding of Battery Materials}
\label{SEC:EXAMPLE3}
%======================================
\ac{SEAMM} is not only a productivity software for reproducible \ac{CMS} workflows and a powerful engine for academic research, but it is also a valuable tool for industrial research. In this example, we highlight a real-world application of \ac{SEAMM} in predicting the transport properties of lithium ion battery materials via \ac{MD} simulations. This study has two main parts. First, we will discuss the simulation of transport properties such as ionic diffusivity, conductivity, and transference number in liquid electrolytes such as EC:EMC (3:7 w/w) and EC:DMC (1:1 w/w) with \ce{LiPF6} at room temperature. Second, we will discuss the thermal expansion in battery electrodes such as \ce{LiCoO2} at finite temperature and their thermal conductivity as a function of the \ac{SOL} at room temperature. 

The performance of lithium-ion batteries at low temperatures or under rapid charge-discharge conditions is governed by the intrinsic transport properties of their liquid electrolytes. Despite their importance, even for conventional electrolytes, the available transport measurement data at various concentrations and temperatures are scarce in the literature. An alternative to measurement techniques is to simulate the system in order to predict the desired transport properties under various operating conditions that are not otherwise easily accessible through experiments. Here, we demonstrate the application of classical \ac{MD} via the \lammps\ code to model transport in EC:EMC (3:7 w/w) and EC:DMC (1:1 w/w) electrolytes with various concentrations of \ce{LiPF6} salt at room temperature. The simulations were performed using \ac{SEAMM} flowcharts to build the \ac{MD} input files, call the \lammps\ calculator, and finally use custom Python scripts for post-processing of the trajectory data. 

\begin{table}
\centering
\begin{tabular}{c | *{3}{c}}
\hline\hline
\textbf{Solvent/Salt} & \textbf{SMILES} & \textbf{Exp. Density ($\gcmcub$)} & \textbf{MD Density ($\gcmcub$)} \\
\hline
\textbf{EC}  & C1COC(=O)O1  & 1.32 & 1.33  \\
\textbf{EMC}  & CCOC(=O)OC  & 1.01 & 1.03  \\
\textbf{DMC}  & COC(=O)OC  & 1.07 & 1.08  \\
\textbf{\ce{LiPF6}} & [\ce{Li+}].F[\ce{P-}](F)(F)(F)(F)F & -- & -- \\
 
 \hline\hline
 \end{tabular}
\caption{Solvent and salt \ac{SMILES} strings, experimental density and MD simulation density at $25^{\circ}$ C}
\label{table:3}
\end{table}

Table \ref{table:3} shows the \ac{SMILES} of different solvents and the salt used in our simulations which are ethylene carbonate (EC), ethyl methyl carbonate (EMC), dimethyl carbonate (DMC), and lithium hexafluorophosphate (\ce{LiPF6}) salt. The initial \ac{MD} configuration is generated using the PACKMOL step, which reads the \ac{SMILES} strings of the solvents and salt based on the given electrolyte formulation and creates a cubic periodic cell to initiate the \lammps\ simulation. The \ac{OPLS-AA} force field \cite{jorgensen1996development, kaminski2001evaluation} were used to describe the bonded and nonbonded parameters of each solvent. The parameters used for the \ce{Li+} cations and \ce{LiPF6} anions were from Jensen \etal\cite{jensen2006halide} and Lopes \etal,\cite{canongia2004molecular} respectively. 

The individual density of each solvent was first validated through \ac{MD} simulations, as shown in Table \ref{table:3}. These densities were used to make an initial guess for the density and hence dimensions of the initial configuration of the electrolyte formulations. The \ac{LAMMPS} simulation involved a structural minimization followed by 2 ns NPT canonical dynamics using a Nose-Hoover thermostat and barostat at $25^{\circ}$ C and 1 atm to equilibrate the system. The cell was then adjusted isotropically to the equilibrium density and the results of 5 production runs of 200 ps of NVT dynamics were averaged. Using \ac{SEAMM}'s \texttt{Diffusivity} module, the position and velocity trajectories were analyzed and the diffusion constant was calculated using the \ac{MSD} approach. The slope of the linear regime in the \ac{MSD} was obtained for each simulation of 200 ps duration and averaged over the total of 1 ns of the production runs to obtain the diffusion constant, defined as
\begin{equation}\label{eq1}
    D (T) = \frac{\langle \text{MSD} (T)\rangle}{6\Delta t} \\
\end{equation}

The electrolyte salt diffusivity, $D_\text{\ce{LiPF6}}(T)$, was calculated from the individual diffusivities of \ce{Li+} cations and $\ce{PF6-}$ anions in the cell as \cite{bizeray2016resolving}
\begin{equation}\label{eq2}
    D_\text{\ce{LiPF6}} (T)  = \frac{2D_\text{\ce{Li+}}(T)~ D_\text{\ce{PF6-}}(T)}{D_\text{\ce{Li+}}(T)+D_\text{\ce{PF6-}}(T)}.
\end{equation}
The electrolyte ionic conductivity, $\sigma(T)$, can also be calculated using the Nernst-Einstein equation,\cite{france2019correlations}
\begin{equation} \label{eq3}
    \sigma (T) = \frac{N_\text{\ce{Li}} e^2}{Vk_BT}(D_{\ce{Li+}}(T)+D_{\ce{PF6-}}(T)),
\end{equation}
where, $N_\text{\ce{Li}}$ is the number of \ce{Li+} ions and $V$ is the volume of the simulation cell, $e$ is the electronic charge, $T$ is the temperature, $D_\text{\ce{Li+}}$ is the \ce{Li+} cation diffusion constant, and $D_\text{\ce{PF6-}}$ is the \ce{PF6-} anion diffusion constant. The cationic transference is expressed in terms of the ratios of the ionic diffusivities,\cite{pesko2017negative}
\begin{equation} \label{eq4}
    t_+  = \frac{D_\text{\ce{Li+}}}{D_\text{\ce{Li+}}+D_\text{\ce{PF6-}}}
\end{equation}

\begin{figure}[!tbph]
	\centering
    % \vspace{-2cm}
	\includegraphics[scale=.6]{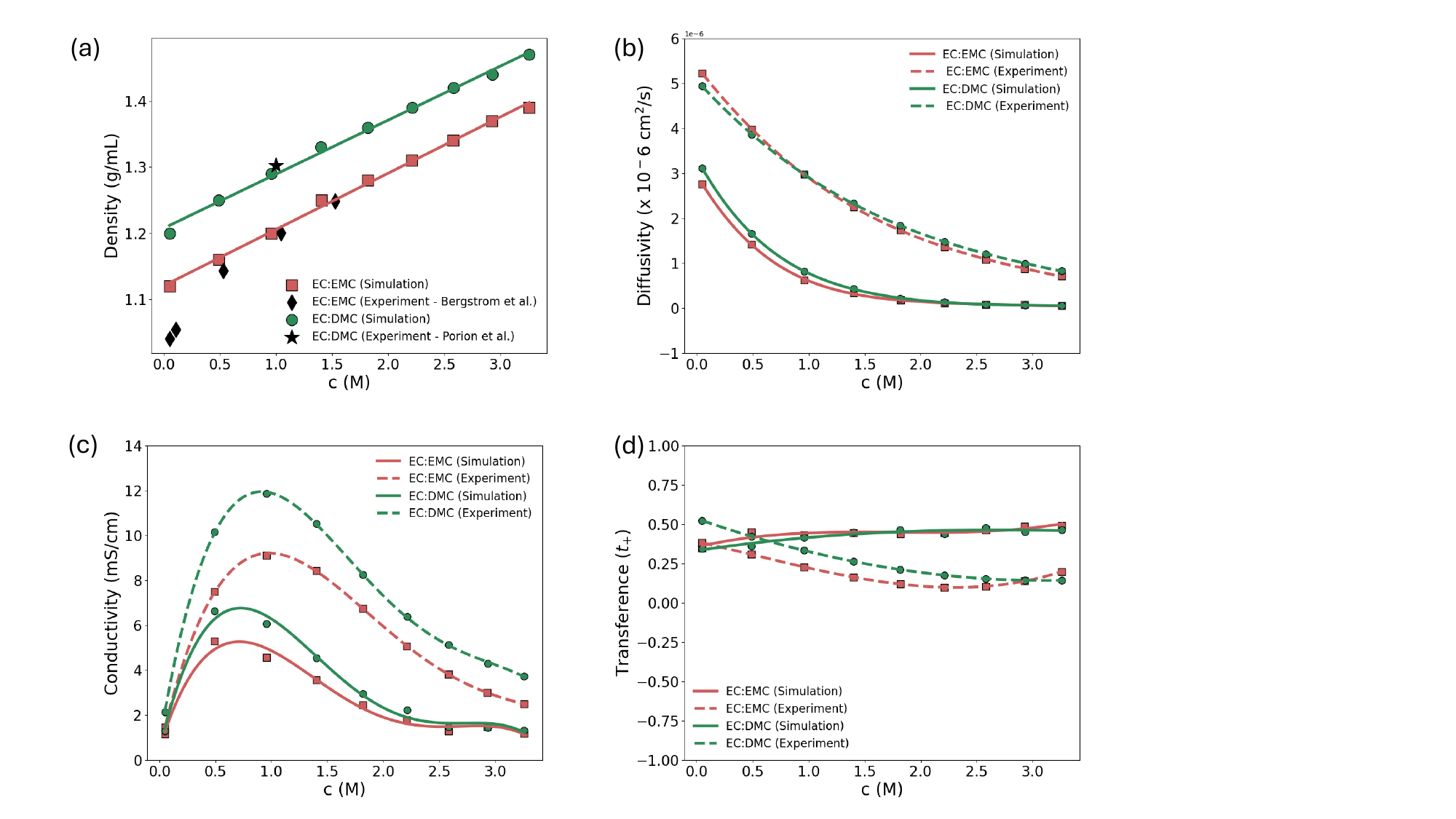}
	\caption{Simulation of transport properties of lithium ion battery liquid electrolytes, EC:EMC (3:7 w/w) and EC:DMC (1:1 w/w), at various molar concentrations ($c$) of \ce{LiPF6} salt at $25^{\circ}$ C. Panels (a-d) show the MD density, ionic diffusivity, ionic conductivity, and cationic transference number, respectively, which are obtained from MD simulation and compared with experimental results performed at $25^{\circ}$ C. The experimental densities in (a) were taken from Refs.~\citenum{bergstrom2021interfacial} and \citenum{porion2013comparative}. The experimental results for diffusivity, conductivity and transference numbers in (b-d) were taken from Ref.~\citenum{landesfeind2019temperature}}
	\label{fig:electrolyte simulation}
\end{figure}

Figure \ref{fig:electrolyte simulation} shows the calculated transport properties from \ac{MD} simulations of EC:EMC (3:7 w/w) and EC:DMC (1:1 w/w) electrolytes, at various molar concentrations ($c$) of \ce{LiPF6} salt at $25^{\circ}$ C. The equilibrium densities [Figure \ref{fig:electrolyte simulation} (a)], obtained from the \ac{MD} simulation, are in agreement with the experimental mass densities of EC:EMC (3:7 w/w) \cite{bergstrom2021interfacial} and EC:DMC (1:1 w/w) \cite{porion2013comparative} electrolytes, especially at salt concentration $> 0.5$ M. At any salt concentration, the EC:DMC (1:1 w/w) mixture resulted in a higher density values than those of EC:EMC (3:7 w/w). Landesfeind and Gasteiger\cite{landesfeind2019temperature} performed an extensive experimental study on the temperature and concentration dependence of the ionic transport properties of these electrolytes. We compare the simulated ionic diffusivity, conductivity and transference number with their result as shown in Figure \ref{fig:electrolyte simulation}(b-d). We found that the predicted ionic transport properties from the simulation underestimated the experimental results, which could be attributed to the default ionic charges of the \ac{OPLS-AA} force-fields. It has been reported in earlier literature that the transport properties in high-concentration electrolytes can be significantly underestimated if the ionic charge correction is neglected \cite{zhou2024optimization, ruza2025benchmarking} in the \ac{MD} simulations conducted with \ac{OPLS-AA} force-fields. Even using the recently improved OPLS4 force-field parameters,\cite{lu2021opls4}  the predicted values of transport and thermodynamic properties of various ions and solvated molecules of common ionic liquids for a wide range of temperature \cite{dajnowicz2022high} were underestimated. However, even though the relative values of the ionic properties are underestimated in the simulations, the overall trends in the experimental property values between electrolyte formulations are accurately conserved within the \ac{MD} simulations. Both experiment and simulation results demonstrate that both electrolyte formulations in our study have similar ionic diffusions. Nonetheless, the EC:DMC (1:1 w/w) mixture exhibited much higher ionic conductivity and cationic transference number than those of EC:EMC (3:7 w/w), due to its higher mass density.

\begin{figure}[!tbph]
	\centering
	\includegraphics{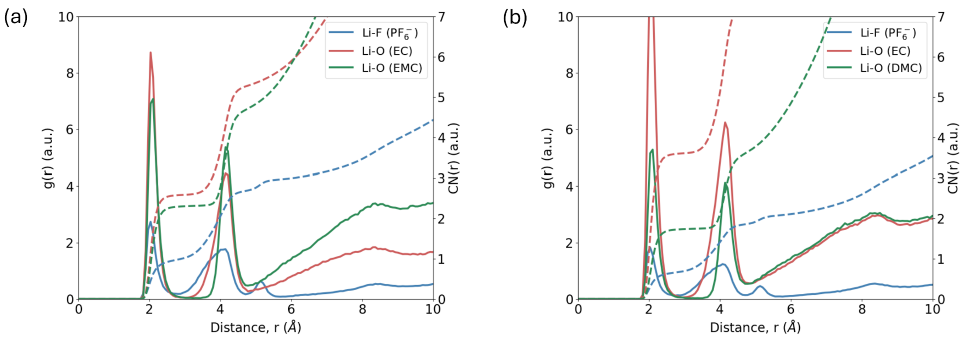}
    % \vspace{-5cm}
	\caption{Calculated radial distribution function, $g(r)$, and corresponding coordination number integrals, $\text{CN}(r)$, of (a) \ce{Li-F} (\ce{PF6-}), \ce{Li-O} (EC), \ce{Li-O} (EMC) in EC:EMC (3:7 w/w) with 1M \ce{LiPF6}, and (b) \ce{Li-F} (\ce{PF6-}), \ce{Li-O} (EC), \ce{Li-O} (DMC) in EC:DMC (1:1 w/w) with 1M \ce{LiPF6}}
	\label{fig:rdf cn}
\end{figure}

One of the key functionalities of \ac{SEAMM} workflows is the ability to write custom python scripts within flowcharts, which enable on-the-fly post-processing of the \ac{MD} simulation data. Figure \ref {fig:rdf cn} shows one such example where the radial distribution function, $g(r)$, and the corresponding coordination number, $\text{CN}(r)$, of solvated \ce{Li+} ions in the above electrolyte formulations were calculated using a custom script that analyzed the \ac{MD} trajectory data. In both formulations, we found that the first solvation shell is mostly occupied by the EC solvent molecules with a most probable distance of 2.05 \AA to \ce{Li+}, which is in agreement with previous \ac{MD} studies of similar electrolyte systems.\cite{hou2019influence} The presence of EMC and DMC molecules within the first solvation shell depends on their relative ratios to the EC solvent molecules. We found a substantial presence of EMC in EC:EMC (3:7 w/w) while the presence of DMC was almost half of EC in EC:DMC (1:1 w/w). The average \ce{Li+} ion coordination number with solvent and salt is 5.84 for EC:EMC (3:7 w/w) and 5.98 for EC:DMC (1:1 w/w), which are also in good agreement with earlier \ac{MD} studies.\cite{hou2019influence} 

\begin{table}
\centering
\begin{tabular}{cc}
\hline\hline
\textbf{Lattice Constant} & \textbf{Expansion (K$^{-1}$)} \\

\hline
\textbf{a}  & 8.36 $\cdot\ 10^{-6}$  \\
\textbf{b}  & 8.36 $\cdot\ 10^{-6}$  \\
\textbf{c}  & 2.04 $\cdot\ 10^{-5}$  \\
 \hline
 \textbf{Average Expansion} & 1.24 $\cdot\ 10^{-5}$ \\
  \hline
  \textbf{Experimental Expansion} & 1.30 $\cdot\ 10^{-5}$ \\
  \hline\hline
 \end{tabular}
\caption{Thermal expansion in \ce{LiCoO2}}

\label{table:4}
\end{table}

\begin{figure}[!tbph]
	\centering
	\includegraphics[width=1.0\textwidth]{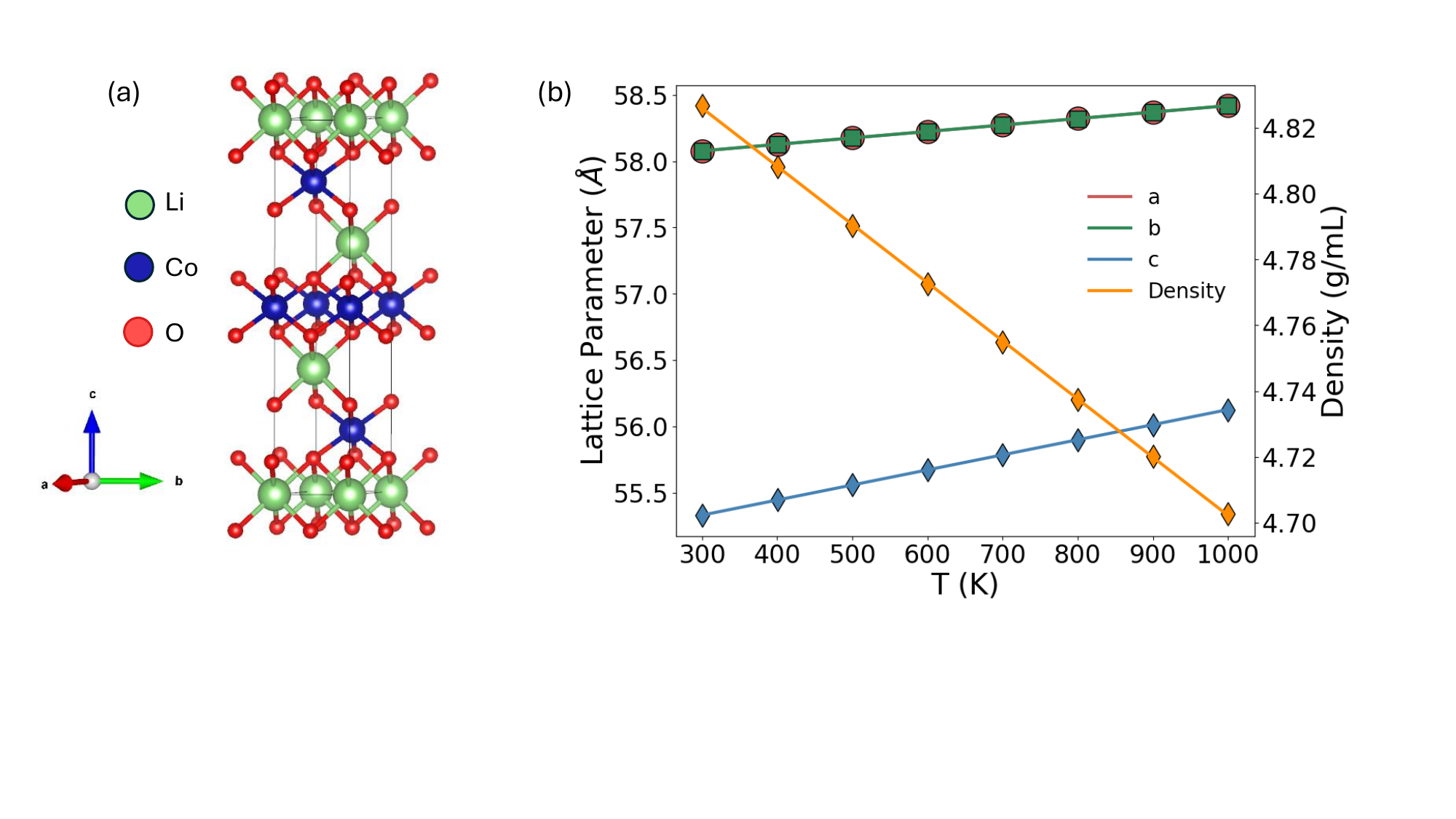}
    \vspace{-3cm}
	\caption{Thermal expansion in \ce{LiCoO2}. (a) The experimental layered (R3m) crystal structure of \ce{LiCoO2} used in the \ac{MD} simulation. The Li, Co and O atoms are represented as green, blue and red spheres. (b) Lattice parameter expansion and change in density as a function of temperature in \ce{LiCoO2}.}
	\label{fig:lco expansion}
\end{figure}

\begin{sidewaysfigure}
	\centering
    % \vspace{-3cm}
	\includegraphics[width=1.\textheight,height=.74\textwidth]{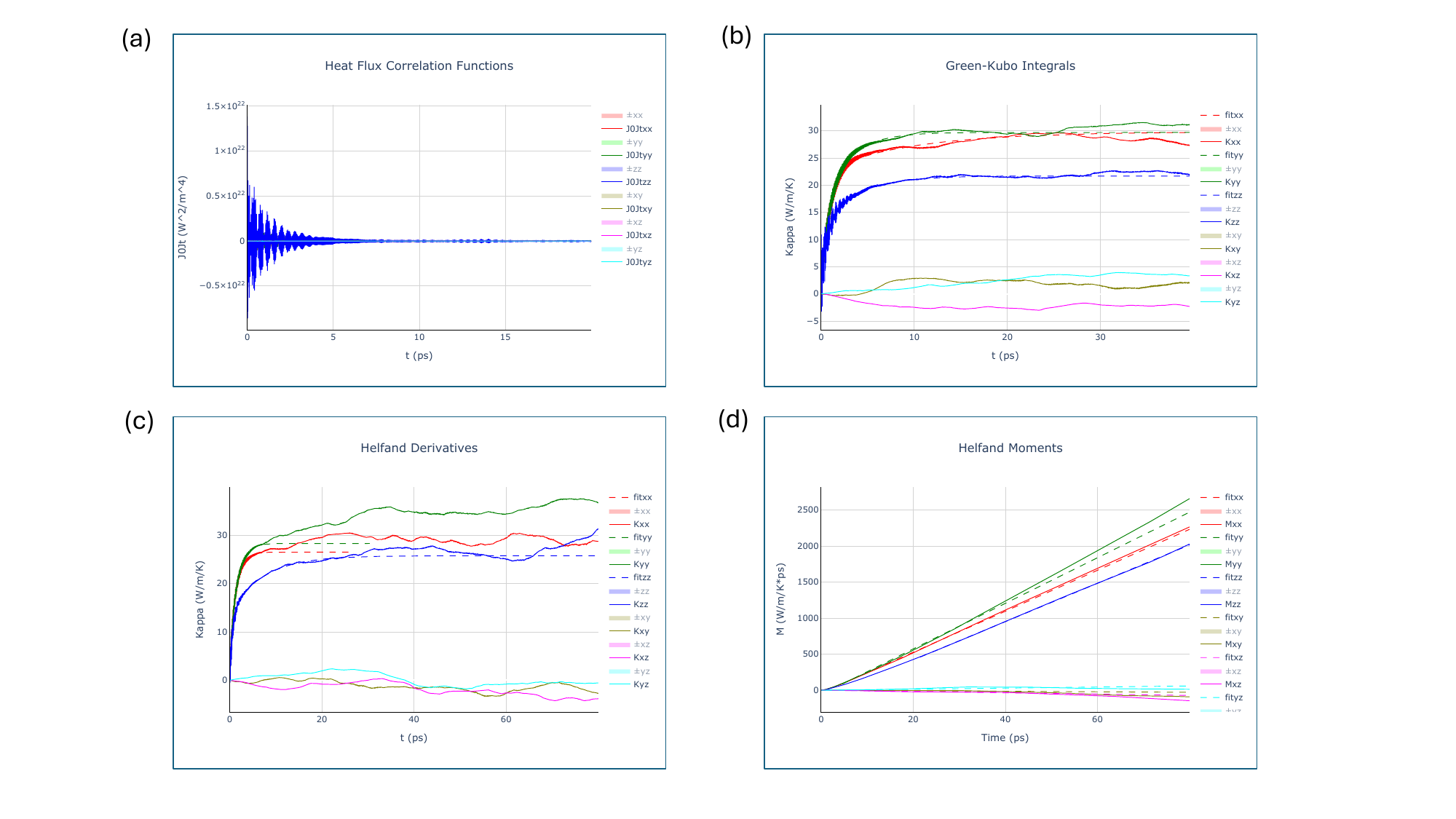}
    % \vspace{-2cm}
	\caption{Thermal conductivity simulation of \ce{LiCoO2} using equilibrium \ac{MD} at $T=298K$. Sample calculation of (a) heat flux auto-correlation function, (b) thermal conductivity fittings of Green-Kubo integrals, (c) thermal conductivity fitting of  Helfand derivatives, and (d) Helfand moments as a function of time in ps.}
	\label{fig:lco tc md}
\end{sidewaysfigure}

\begin{figure}[!tbph]
	\centering
	\includegraphics{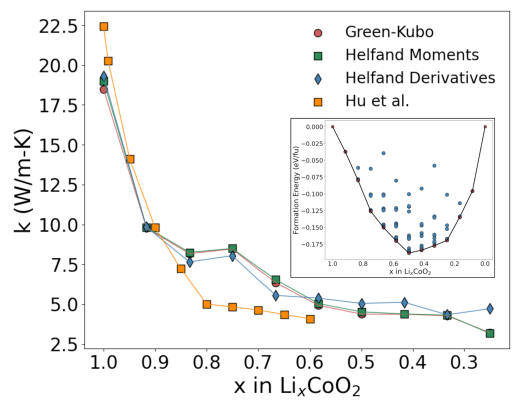}
	\caption{Calculated thermal conductivity of Li$_x$CoO$_2$ as a function of Li concentrations at $T=298K$ from the \ac{MD} simulation. The inset shows a  convex hull diagram calculated using ab initio ground state formation energies to predict the lowest energy Li vacancy structures. }
	\label{fig:lco tc}
\end{figure}

In the next part of our study, we focus on the prediction of the thermal expansion of \ce{LiCoO2} at finite temperature using \lammps, as well as the simulation of thermal conductivity at various \ac{SOL} at room temperature. We used \ac{SEAMM} flowcharts for each step of the equilibrium \ac{MD} simulations. Crystalline layered \ce{LiCoO2} with rhombohedral symmetry (space group $R3m$, as shown in Figure \ref{fig:lco expansion}(a)\cite{rodriguez2003quantitative}), and the Buckingham potential\cite{buckingham:1938:264} developed by Fisher \etal\cite{aj2010atomic} were used for the simulations. The potential energy between pairs of ions in each crystalline solid was calculated by combining the long-range Coulombic component with a short range repulsive and attractive van der Waals interaction as described in Ref.~\citenum{aj2010atomic}. The short-range interactions, $\phi_{i,j}$, were described by a Buckingham potential of the form \cite{aj2010atomic}
\begin{equation} \label{eq5}
    \phi_{i,j} (r_{ij}) = A_{ij} \exp(-r_{ij}/\rho_{ij})-C_{ij}/r_{ij}^6,
\end{equation}
where, $r_{ij}$ is the distance between the ions $i$ and $j$, $A_{ij}$, $\rho_{ij}$ and $C_{ij}$ are parameters fit to experimental lattice parameters and ion positions as described in Ref.~\citenum{aj2010atomic}. A cutoff distance of 14 \AA  was used for the Coulombic and Buckingham terms. First, equilibrium \ac{MD} simulations with the isothermal-isobaric (NPT) ensemble were performed using a Nose-Hoover thermostat and barostat at a temperature $T$ and a pressure of 1 atm for 200 ps, using a time step of 1 fs, to achieve thermal equilibration and to relax the structure. The lattice parameters of the relaxed geometry at each temperature T were then collected for the expansion calculation. 

Figure \ref{fig:lco expansion}(b) shows the in-plane and out-of-plane lattice parameters and the change in mass density in \ce{LiCoO2} as a function of temperature. The corresponding expansion coefficients (defined as $\Delta L/L \Delta T$, where $L$ is the length of a lattice constant) along the lattice direction of $a$, $b$ and $c$ are shown in Table 4. The in-plane expansions are identical for the $a$ and $b$ directions and much smaller than the expansion along the out-of-plane $c$ direction. This anisotropic lattice expansion is expected since the \ce{LiCoO2} has a layered crystal structure where the \ce{Li} and transition metal layers interact with each other with van der Waals interactions. The average thermal expansion coefficient predicted by the MD simulation ($1.24 \cdot 10^{-5}$ K$^{-1}$) is in close agreement with the experimental expansion coefficient from the polycrystalline \ce{LiCoO2}  pellets ($1.3 \cdot 10^{-5}$ K$^{-1}$) \cite{cheng2017elastic}.  

Equilibrium \ac{MD} methods such as Green-Kubo formalism,\cite{green1954markoff, kubo1957statistical} Helfand derivatives, and Helfand moments\cite{helfand1960transport, gaspard2005chaos, viscardy2007transport} were used to predict the thermal conductivity of \ce{LiCoO2}. According to the Green-Kubo theory, which is derived from the fluctuation dissipation theorem and linear response theory, the thermal conductivity tensor $\kappa^{\alpha\beta}$ is proportional to the time integral of the heat flux autocorrelation function \cite{green1954markoff, kubo1957statistical},

\begin{equation} \label{eq6}
    \kappa^{\alpha\beta} (T) = \frac{1}{k_{B}T^2V} \ \int_{0}^{\infty} \langle J^{\alpha}(0) \cdot J^{\beta}(t) \rangle\ dt,
\end{equation}
where, $J^{\alpha} (t)$ is the $\alpha$th Cartesian component of the time-dependent heat current, $k_B$ is the Boltzmann constant, $V$ is the system volume, $T$ is the system temperature, and $\langle \cdot \rangle$ denotes the ensemble average. In addition to the Green-Kubo approach, the Helfand moment and Helfand derivative approaches, which follow directly from the Green-Kubo expression\cite{helfand1960transport, gaspard2005chaos, viscardy2007transport}, were used to calculate the thermal conductivity. The Helfand moment $G^{\alpha}(t)$ is defined in terms of the time integral of the heat current component $J^{\alpha}$ as
\begin{equation} \label{eq7}
    G^{\alpha}(t) = G^{\alpha}(0) + \int_{0}^{t} J^{\alpha}(\tau)\  d\tau
\end{equation}
in which, the ensemble-averaged quantity, $M^{\alpha\beta}$, is defined as \cite{gaspard2005chaos}
\begin{equation} \label{eq8}
    M^{\alpha\beta}(t) = \langle (G^{\alpha}(t)-G^{\alpha}(0)) (G^{\beta}(t)-G^{\beta}(0)) \rangle.
\end{equation}
For a sufficiently long simulation time, $M^{\alpha\beta}(t)$ grows linearly as a function of time with a slope that is directly proportional to the corresponding thermal conductivity tensor component $\kappa^{\alpha\beta} (T)$.\cite{pereverzev2022heat} Therefore, the thermal conductivity tensor $\kappa^{\alpha\beta} (T)$ is then defined in the Helfand approach as
\begin{equation} \label{eq9}
    \kappa^{\alpha\beta} (T) = \frac{1}{2tk_{B}T^2V} \ \lim_{t\to\infty} M^{\alpha\beta}(t)
\end{equation}

The advantage of the Helfand method over the more traditional Green-Kubo approach is two-fold. First, its convergence with simulation time is superior to that of Green-Kubo. Second, it is easier to fit the straight line given by Equation \ref{eq9} than the exponential rise of the Green-Kubo function, which is complicated by the increasing noise as a function of time.

In order to calculate the time-dependent heat current autocorrelation function and apply it to the thermal conductivity calculation using the Green-Kubo and Helfand approaches, we performed production \ac{MD} simulation with various run times. Before running the production runs, the system was further equilibrated for 200 ps, with a time step of 1 fs, of canonical (NVT) dynamics at a temperature $T$ using a Nose-Hoover thermostat. Then simulations using the microcanonical (NVE) ensemble with a run length of 200 ps, 400 ps and 800 ps were carried out 10 times for each run length, for a total of 30 production runs, to calculate the heat flux autocorrelation function. The results of all the production runs were averaged to give the final thermal conductivity. Figure \ref{fig:lco tc md} shows a sample calculation of heat flux autocorrelation function, the Green-Kubo integrals, Helfand derivative, and Helfand moments for fully lithiated \ce{LiCoO2} at $T=300K$ for a 400 ps production run.

Figure \ref{fig:lco tc} shows the calculated thermal conductivity of \ce{LiCoO2} at various lithium concentrations at temperature of $T=298 K$. Each \ce{Li} vacancy-ordered structure was first generated using \ac{DFT} to calculate the formation energy of the convex hull of the ground state ($T=0K$). The inset in Figure \ref{fig:lco tc} shows the convex hull diagram where the lowest energy structure was considered to be the most stable structure at any \ce{Li} concentration $x$. Once these structures were identified, we used the classical \ac{MD} simulation described above to calculate the thermal conductivity at $T=298 K$. All three approaches (Green-Kubo, Helfand moments, and Helfand derivatives) resulted in very similar thermal conductivities at any $x$ considered in our calculations. Our results indicate that with the decrease in \ce{Li} concentration during battery charging, the thermal conductivity decreases drastically from about 20 W/m/K at $x=1$ to 3.2 W/m/K at $x=0.3$. These results agree quantitatively with an earlier \ac{MD} study by Hu \etal\cite{hu2018significant} where the nonequilibrium \ac{MD} was used for the prediction of the thermal conductivity in \ce{LiCoO2}.

%======================================
\section{Conclusions}\label{SEC:CONCLUSION}
%======================================
We have presented SEAMM, a new open source productivity tool and workflow
environment for computational modeling and simulations of molecular and
solid-state materials. SEAMM is underpinned by an extensive data model for
molecular and crystalline structures implemented in an in-memory relational
database using a traditional schema. The database also stores calculated
properties and results by describing them with metadata using a star schema to
connect the data to the metadata, allowing the data model to be modified and
extended on the fly. This flexibility will permit SEAMM to adapt and change as
current simulation tools evolve and new tools and methods are developed. An
object-oriented facade is layered on top of the database, giving SEAMM a
conventional data model from the software developers point of view. However, the
flexibility and extensibility of the underlying model is carried through to the
object-oriented model.

The heart of SEAMM are flowcharts, which represent simulation workflows in a
graphical form. The functionality that users see in SEAMM is provided by
independent software modules that plug-in to SEAMM and share data through the
in-memory database. Each plug-in corresponds to a step in a flowchart, presents
its own graphical interface to users, implements the needed computation or
simulation, and analyzes and presents the results. The only connection between
plug-ins is the underlying data model in SEAMM. This decentralized approach
allows any number of different groups to develop plug-ins with little or no
coordination between groups. Users then create flowcharts to implement their
workflows, combining plug-ins to do the simulations and modeling that are needed
for their scientific or engineering problem.

There are a number of scriptable workflow tools for computational molecular and
materials science (CMS), as well as a number of graphical tools for specific
tasks. However, there is no general open source graphical workflow system for
CMS. SEAMM fills this gap and extends the range of CMS tools. With a focus on
productivity, on reducing the entry barrier to using a wide range of CMS
simulation tools, and on the reproducibility and transparency of the workflows,
SEAMM is a valuable contribution to the community.

\begin{acknowledgement}
The present work is funded by the National Science Foundation grants OAC-1547580 and CHE-2136142. PS and MM also acknowledge the Advanced Research Computing (\url{https://arc.vt.edu}) at Virginia Tech for providing computational resources and technical support that have contributed to the results reported within this manuscript.
\end{acknowledgement}

%%%%%%%%%%%%%%%%%%%%%%%%%%%%%%%%%%%%%%%%%%%%%%%%%%%%%%%%%%%%%%%%%%%%%
%% Acronyms (alphabetic order)
%%%%%%%%%%%%%%%%%%%%%%%%%%%%%%%%%%%%%%%%%%%%%%%%%%%%%%%%%%%%%%%%%%%%%

\begin{acronym}
    \acro{API}{Application Programming Interface}
    \acro{AQME}{automated quantum mechanical environments for researchers and educators}
    \acro{ASE}{Atomic Simulation Environment}
    \acro{ACF}{autocorrelation function}
    \acro{BSE}{Basis Set Exchange}
    \acro{CI-NEB}{climbing image nudged elastic band}
    \acro{CMS}{computational molecular science}
    \acro{DOI}{digital object identifier}
    \acro{DFT}{density functional theory}
    \acro{GUI}{graphical user interface}
    \acro{IDPP}{image dependent pair potential}
    \acro{JSON}{JavaScript Object Notation}
    \acro{LAMMPS}{Large-scale Atomic/Molecular Massively Parallel Simulator}
    \acro{MD}{molecular dynamics}
    \acro{MSD}{Mean Square Displacement}
    \acro{MVC}{model-view-controller}
    \acro{NEB}{nudged elastic band}
    \acro{OPLS-AA}{Optimized Potentials for Liquid Simulations --- All Atom}
    \acro{PyPI}{the Python Package Index}
    \acro{RDBMS}{Relational Database Management System}
    \acro{REST}{Representational State Transfer}
    \acro{RI}{resolution of identity}
    \acro{SEAMM}{Simulation Environment for Atomistic and Molecular Modeling}
    \acro{SDF} {structure-data format}
    \acro{SMILES}{Simplified Molecular Input Line Entry System}
    \acro{SOL}{state of lithiation}
    \acro{SQL} {structured query language}
    \acro{VASP}{the Vienna Ab initio Simulation Package}
\end{acronym}

% \begin{suppinfo}

%   % A listing of the contents of each file supplied as Supporting Information
%   % should be included. For instructions on what should be included in the
%   % Supporting Information as well as how to prepare this material for
%   % publications, refer to the journal's Instructions for Authors.

%   % The following files are available free of charge.
%   % \begin{itemize}
%   %   \item Filename: brief description
%   %   \item Filename: brief description
%   % \end{itemize}
% \input{seamm/supplement}

% \end{suppinfo}

\bibliography{ms}
\end{document}

%--------------------------------------
\newpage
%--------------------------------------
\tableofcontents
% \setcounter{secnumdepth}{-1}
%--------------------------------------
\newpage
%--------------------------------------
%%%%%%%%%%%%%%%%%%%%%%%%%%%%%%%%%%%%%%%%%%%%%%%%%%%%%%%%%%%%%%%%%%%%%
%% Start the main part of the SI here.
%%%%%%%%%%%%%%%%%%%%%%%%%%%%%%%%%%%%%%%%%%%%%%%%%%%%%%%%%%%%%%%%%%%%%

%======================================
\addcontentsline{toc}{section}{Introduction}
\section{Introduction}\label{SUPPSEC:INTRO}
%======================================
This supplementary material covers the simulations and results of Example 2 in the paper in more detail. The first part describes the calculations underlying Tables 1 and 2. The second part covers the calculations with ReaxFF and TorchaANI, expanding on the results mentioned in the paper and some exploration of the nature of the potential energy surface for these methods.

%======================================
\addcontentsline{toc}{section}{Flowchart for Reaction Paths}
\section{Flowchart for Reaction Paths}
\label{SUPPSEC:FLOWCHART}
%======================================
Finding the transition state for a reaction and calculating the energies and enthalpies of the reactants, transition state, and products requires a number of distinct calculations:
\begin{enumerate}
    \item Determining the structures of the reactant(s) and product(s);
    \item Locating the transition state between the reactants(s) and product(s);
    \item Refining the structure of the transition state;
    \item Calculating the thermochemical functions in the harmonic approximation for the reactant(s), transition state, and product(s)
\end{enumerate}

\begin{figure}[p]
	\centering
	\includegraphics[height=0.9\textheight]{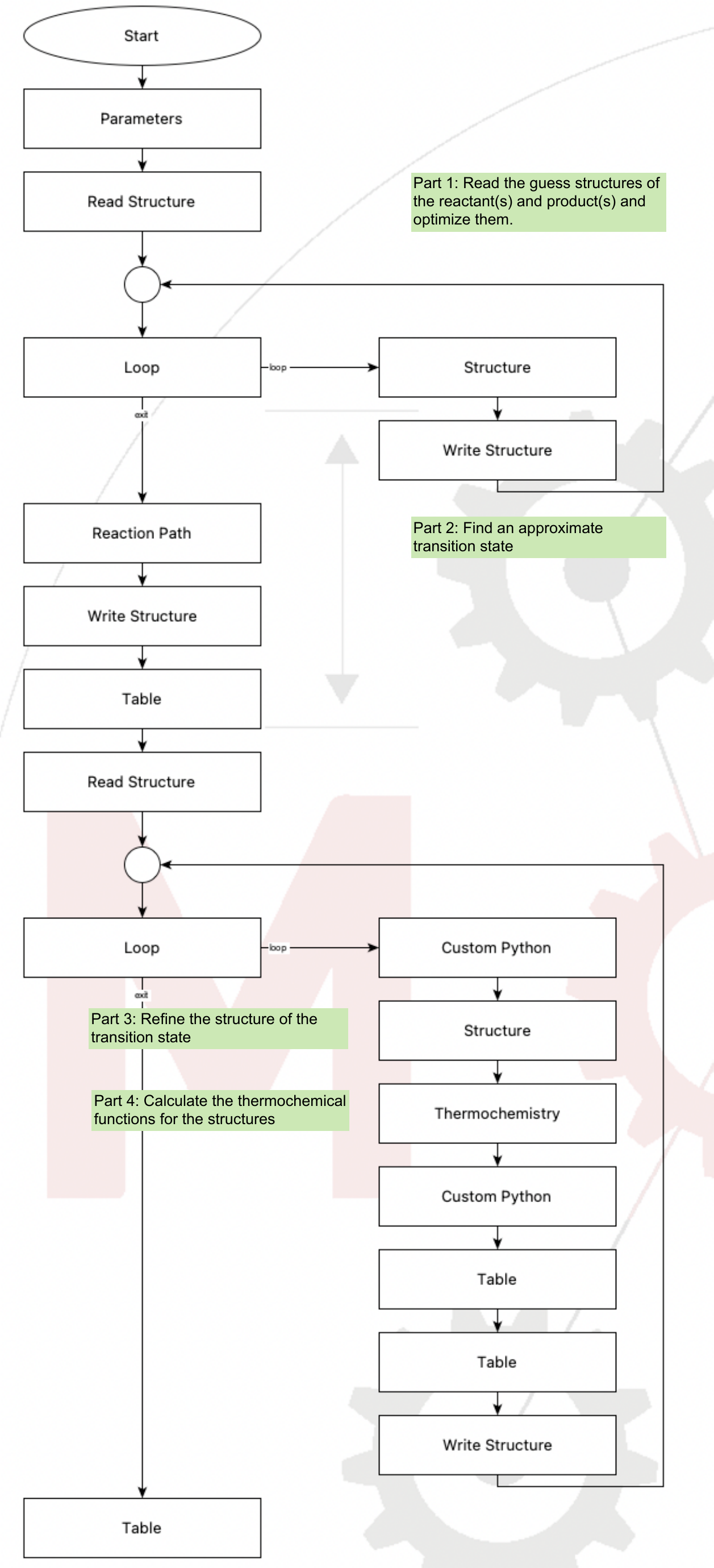}
	\caption{Flowchart for Reaction Thermochemistry}
	\label{fig:reaction flowchart}
\end{figure}

Each of these parts can be handled with a separate flowchart in \ac{SEAMM}, or two or more parts can be combined in a single flowchart. Figure \ref{fig:reaction flowchart} shows a single flowchart for handling all the needed calculations, which is convenient for running either many different methods for a single reaction, or investigating many reactions.

This flowchart corresponds directly to the steps above, as shown by the green text boxes. The next subsections describe the flowchart in some detail because it nicely illustrates the flexibility and power of \ac{SEAMM}. However, since the flowchart implements a reasonably complicated workflow, the description is quite lengthy, and may be skipped if the detail is not of interest.

This flowchart uses loops to handle the initial structures as well as the final structures, \texttt{Table} steps to capture and tabulate the key results, and \texttt{Custom Python} steps to customize small parts of the flowchart. These steps will be covered in more detail below.

%--------------------------------------
\addcontentsline{toc}{subsection}{Initialization}
\subsection{Initialization}
\label{SUPPSUBSEC:INITIALIZATION}
\begin{figure}[!tbph]
	\centering
	\includegraphics[width=\textwidth]{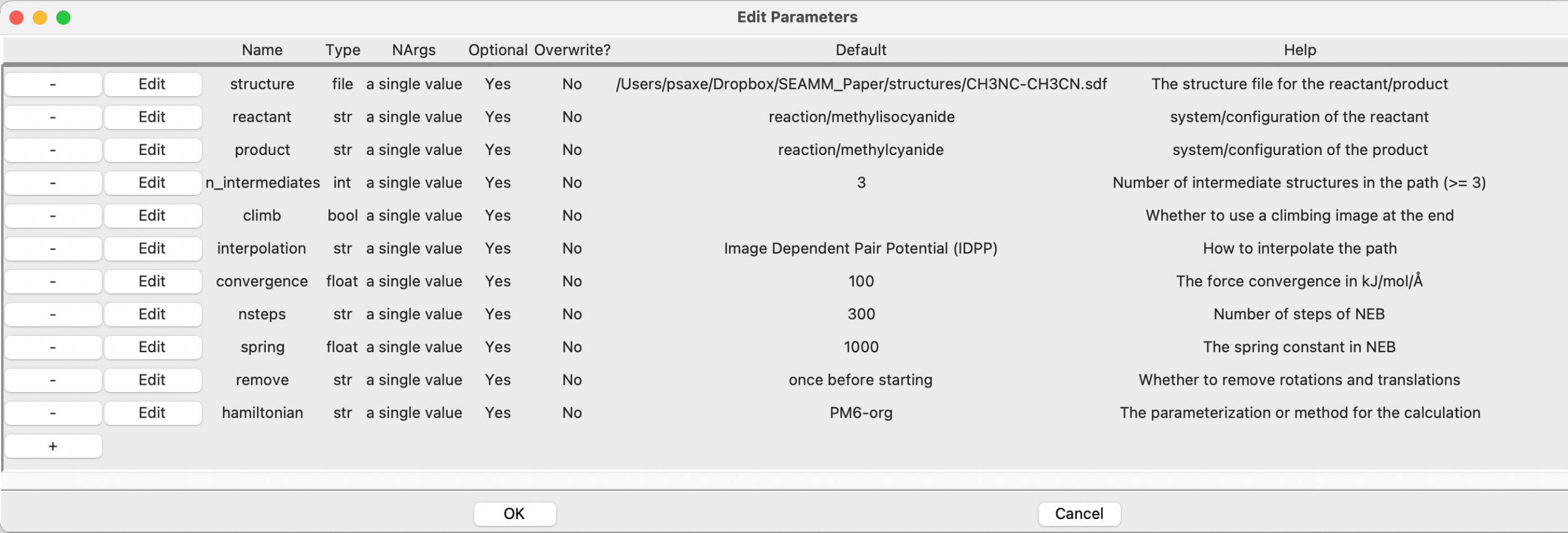}
	\caption{Control parameters for the reaction path flowchart}
	\label{fig:control step}
\end{figure}

%--------------------------------------
 It starts with a \texttt{Parameters} step that defines the value of variables that control parts of the flowchart. As mentioned, this makes the flowchart more general because the user can set the parameters when submitting the job, rather than edit the flowchart. Figure \ref{fig:control step} shows the dialog for editing the control parameters. 

The first three parameters define the reactants and products for the reaction by giving the file containing the structures --- in this case an \ac{SDF} file --- and the names of the structures in that file that are the reactants and products. The assumption is that these structures have been prepared and placed in a file, which gives considerable flexibility in defining the reactant and product systems. Using these variables to define the reactants and products makes the flowchart general and applicable to many different reactions.

The next group of parameters control the \ac{NEB} method\cite{neb:1998,neb:2000,neb:2019} used to find an approximate transition state. The \ac{NEB} employs a number of structures between the reactants and products connected from the reactants through the intermediate structures to the products with an elastic band. \textit{n\_intermediates} defines how many structures to use between the reactants and products, while \textit{spring} gives the force constant for the elastic bands between adjacent structures. \textit{interpolation} determines the initial guess for the intermediate structures, which can currently be linear interpolation or the \ac{IDPP}\cite{idpp:2014} method. \textit{climb} controls whether to use the standard \ac{NEB} method or \ac{CI-NEB}\cite{ci-neb:2000} where the highest energy image is allowed to climb upwards towards the transition state, rather than remaining roughly in the middle between its two adjacent intermediate structures. \textit{remove} controls whether to remove rigid rotations and translations and rotations between the various structures, and if so whether to do so just at the beginning of the calculation or at each step. Finally, \textit{convergence} and \textit{nsteps} give the convergence criterion for the calculation and a maximum number of steps before giving up if the calculation does not converge. 

The last parameter, \textit{hamiltonian}, specifies the parametrization for semiempirical methods, or the method for textit{ab initio} methods, or the forcefield for molecular mechanics. This allows the flowchart to be used for a variety of different approaches without editing. This parameter is used in the all the steps from the initial structure optimization through the reaction path to the final structure optimization and thermochemistry calculation.

Together, these parameters give considerable flexibility in defining the calculations in the flowchart, allowing the user to tune the calculation to their needs.

%--------------------------------------
\addcontentsline{toc}{subsection}{Optimization of the Reactant and Product Structures}
\subsection{Optimization of the Reactant and Product Structures}
\label{SUPPSUBSEC:OPTIMIZATION}
\begin{figure}[!tbph]
	\centering
	\includegraphics[width=\textwidth]{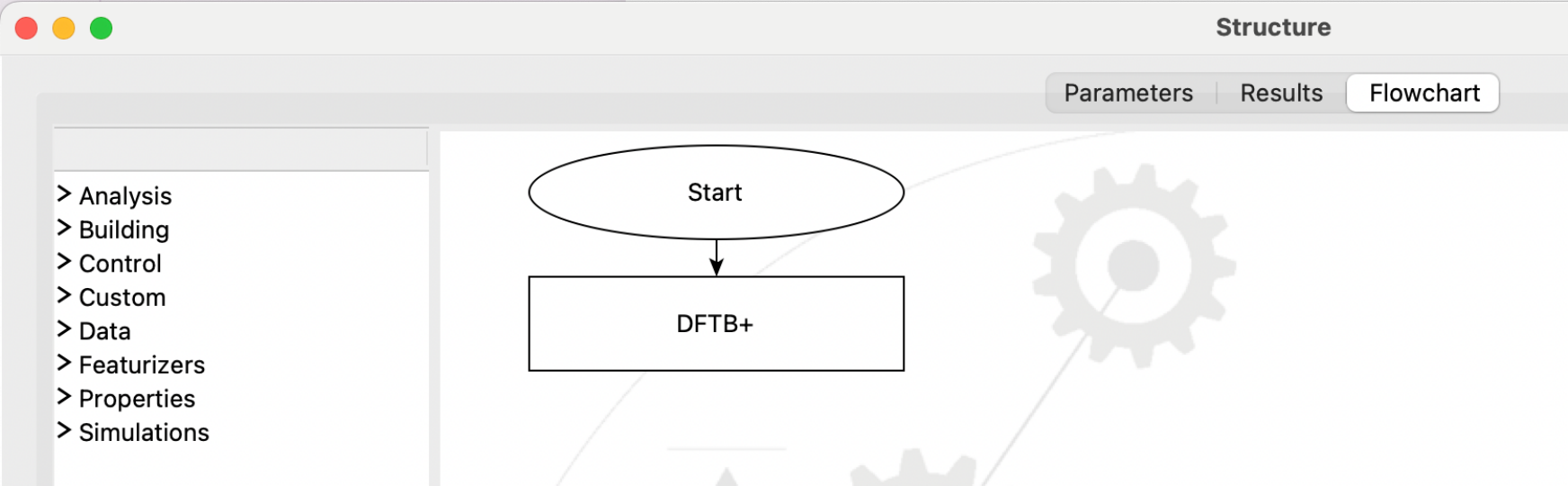}
	\caption{An example of the subflowchart common to the \texttt{Structure}, \texttt{Reaction Path}, \texttt{Thermochemistry}, and \texttt{Energy Scan} steps.}
	\label{fig:subflowchart}
\end{figure}

%--------------------------------------
After reading the structure file, the flowchart continues with a loop over the structures in the file, which are then optimized using the \texttt{Structure} step, which wraps the open source code \texttt{geomeTRIC}\cite{geomeTRIC} to handle the structureal optimization. As has been mentioned, the \texttt{Stucture}, as well as the \texttt{Reaction Path}, and \texttt{Thermochemistry} steps rely on a subflowchart to calculate the energy and forces for the structure, as shown in Figure \ref{fig:subflowchart}, which uses \dftb\ for the energy and forces.

\begin{figure}[!tbph]
	\centering
	\includegraphics[width=\textwidth]{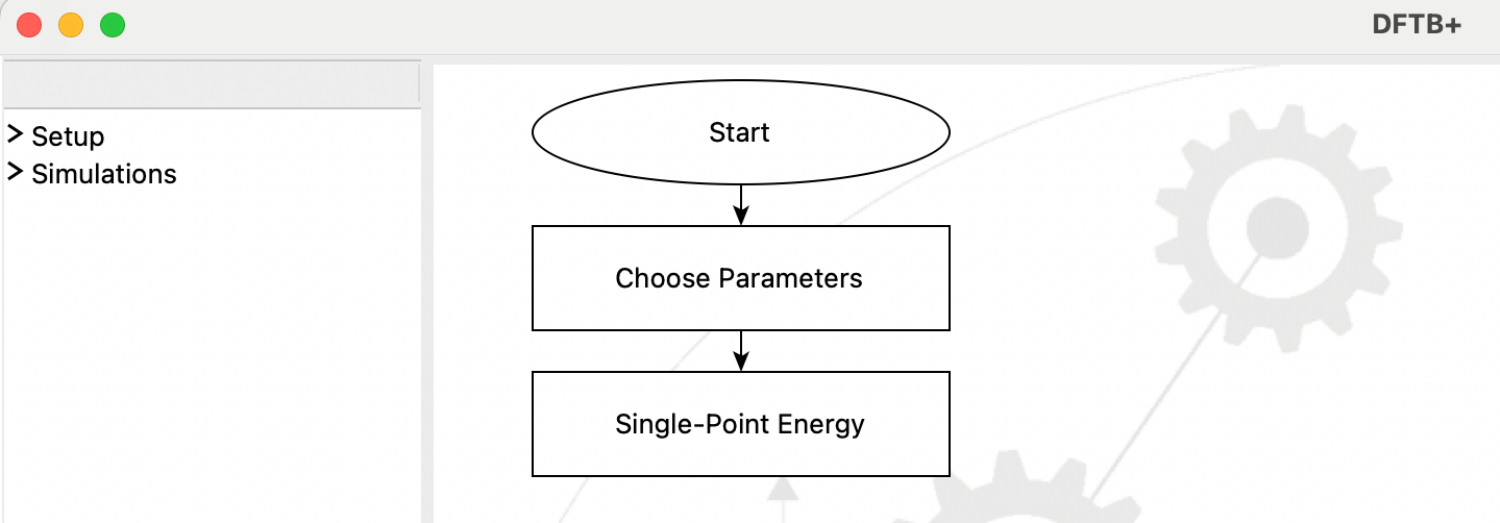}
	\caption{The flowchart to calculate the energy and forces using \dftb.}
	\label{fig:dftb+ flowchart}
\end{figure}

The flowchart to calculate the energy and forces using \dftb\, shown in Figure \ref{fig:dftb+ flowchart} is straightforward. Using any other simulation code to calculate the energy and forces is similarly simple, making it easy to switch between codes.

%--------------------------------------
\addcontentsline{toc}{subsection}{The Reaction Path and Approximate Transition States}
\subsection{The Reaction Path and Approximate Transition States}
\label{SUPPSUBSEC:RXNPATH}
%--------------------------------------
Once the structures for the reactants and products are optimized, the flowchart moves to the \texttt{Reaction Path} step, which uses the \ac{NEB} and \ac{CI-NEB} methods as implemented in \ac{ASE}\cite{Larsen:2017:273002} to find and approximate reaction path and transition state. The \texttt{Reaction Path} step uses an identical subflowchart to that shown above for the \texttt{Structure} step. Indeed, \ac{SEAMM} supports copying and pasting flowcharts so the subflowchart can simply be copied from the \texttt{Structure} step. As noted above, several parameters controlling the \ac{NEB} algorithm are defined by the initial \texttt{Parameters} step as variables, giving considerable flexibility in how to approach the \ac{NEB} calculation.

As the steps in the flowchart run, the various structures that are generated are stored in the internal database in SEAMM for later use. In this flowchart the \texttt{Write Structure} step immediately after the \texttt{Reaction Path} step writes an \ac{SDF} file with all the structures along the reaction path from the \ac{NEB} calculation so that they can be viewed or used later. After creating a table with the \texttt{Table} step, which will be discussed a bit later, the flowchart proceeds to reading the structures for the stationary points that the \texttt{Reaction Path} step found and wrote to a specific \ac{SDF} file. Note that while we have been talking as if there is a single transition state in the reaction path, that is the reaction is an elementary reaction, the \ac{NEB} method is capable of handling more complex reactions with multiple transition states and intermediate structures between them. In such cases, all the stationary point structures would be output from the \texttt{Reaction Path} step and used in the subsequent steps of the flowchart.

%--------------------------------------
\addcontentsline{toc}{subsection}{Optimization of the Key Structures and Their Thermochemistry}
\subsection{Optimization of the Key Structures and Their Thermochemistry}
\label{SUPPSUBSEC:OPTIMIZATION}
%--------------------------------------
The final \texttt{loop} step in the flow chart handles the optimization of the structures of the stationary points along the reaction path. If the reaction is an elementary reaction with just one transition state, there will be three structures: the reactants, the transition state, and the products. For more complicated reactions, there would be additional transition states as well as intermediate minima between them. In all cases, the loop optimizes each structure in the \texttt{Structure} step, calculating the thermochemical functions for each structure in the \text{Thermochemistry} step, which uses the identical subflowchart to the initial \texttt{Structure} step, which can be copied and pasted into this step. Internally the \texttt{Thermochemistry} is using the \texttt{thermochemistry} module from \ac{ASE}\cite{Larsen:2017:273002} to calculate the harmonic approximation to the thermochemical functions of the vibrational frequencies.

\begin{figure}[!tbph]
	\centering
	\includegraphics[width=0.7\textwidth]{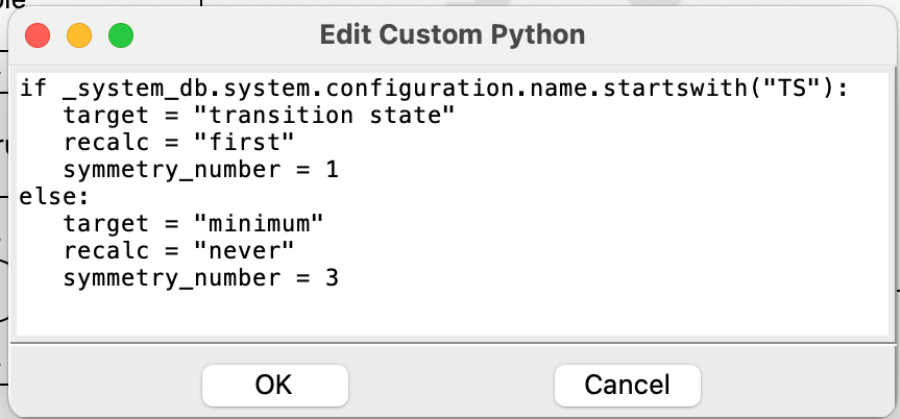}
	\caption{The Python code to set the required parameters for the \texttt{Structure} and \texttt{Thermochemistry} steps.}
	\label{fig:custom step}
\end{figure}

There is a small issue with the optimizer that the \texttt{Custom Step} at the beginning of the loop solves. The optimizer in the \texttt{Structure} step needs to know whether it is targeting a minimum or a transition state. At the moment, the \texttt{Thermochemistry} step requires the symmetry number for the structure. In a future release, the \texttt{Thermochemistry} step will be enhanced to automatically determine the symmetry number, but for the moment it needs to be set manually or using a variable as shown in Figure \ref{fig:custom step}.

The somewhat cryptic first line is accessing the database in \ac{SEAMM} (\_system\_db) and getting the name of the current \textit{system} and \textit{configuration}, which the \texttt{Reaction Step} ensures starts with ``TS'' if it is a transition state. The rest of the Python is self-explanatory. Unfortunately hard-wiring the symmetry number in this way makes the flowchart much less general, and is a temporary fix until the code can correctly determine the symmetry number from the structure.

\begin{figure}[!tbph]
	\centering
	\includegraphics[width=0.7\textwidth]{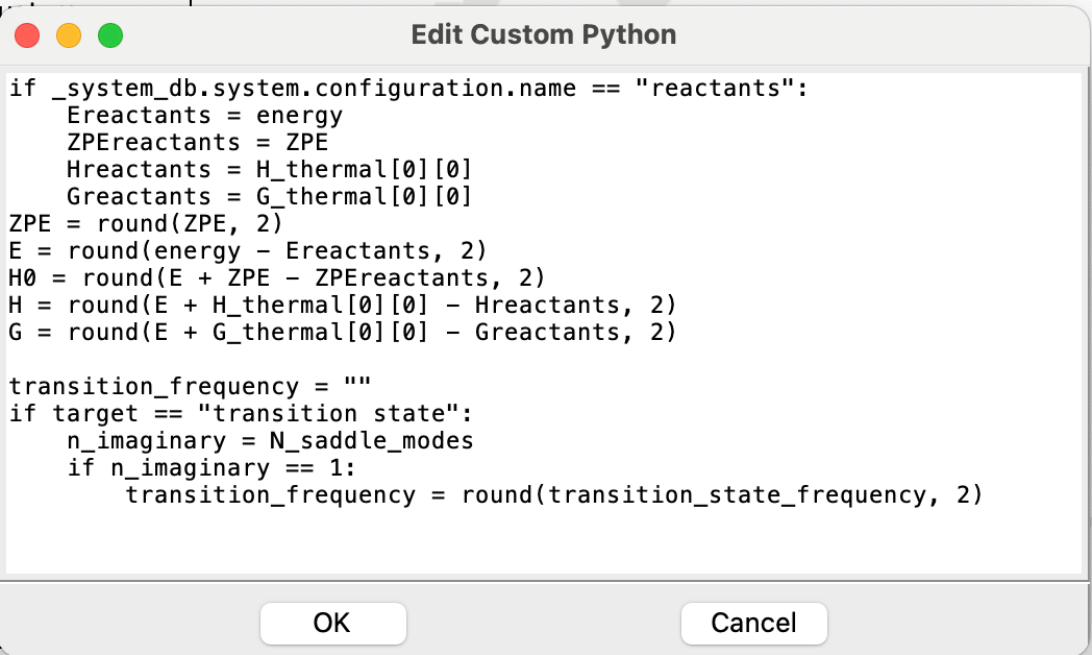}
	\caption{The Python code handle the computed energies and enthalpies.}
	\label{fig:custom step 2}
\end{figure}

We are almost done! The last part of the loop manipulates the energies that the \texttt{Structure} and \texttt{Thermochemistry} saved into variables, which can be accessed by any subsequent step, and then the two \texttt{Table} steps add a row to the table and store it as a \ac{CSV} file to capture the key results in a convenient form for the user. Figure \ref{fig:custom step 2} shows the \texttt{Custom} step that transforms the computed energies and enthalpies into values relative to the reactants, and also captures the imaginary frequency of the transition states. 

\begin{figure}[!tbph]
	\centering
	\includegraphics[width=0.7\textwidth]{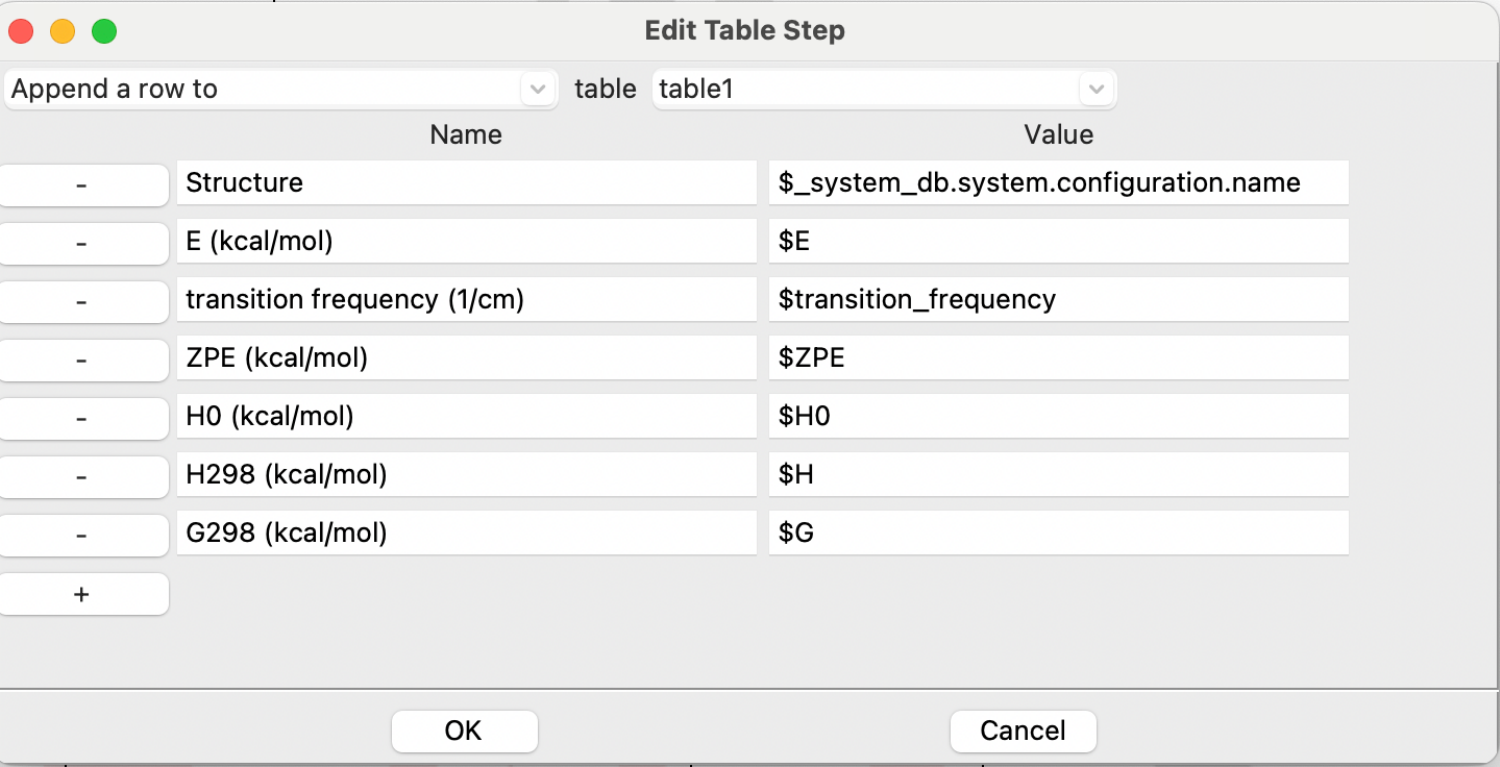}
	\caption{Adding a row of results to the table.}
	\label{fig:table step}
\end{figure}

This is quite straightforward but shows the value of being able to customize the flowchart with a little bit of Python code. In this case, we are able to transform the energies into relative energies, which the next step adds to the table, as shown in Figure \ref{fig:table step}. 

Note that the \texttt{Table} step is using the variables that were just defined in the \texttt{Custom} step shown in Figure \ref{fig:custom step 2}. This illustrates how the Python in \texttt{Custom} steps can access variables set by other steps, and how variables can be used in any of the fields in dialogs if preceded by a dollar sign (`\$') to indicate that it is a variable rather than a literal value. This ability to define and use variables, and insert small fragments of custom code, greatly enhance the capabilities of SEAMM and its flowcharts.

%======================================
\addcontentsline{toc}{section}{Reaction Path for the Rearrangement of Methylisocyanide to Isonitrile}
\section{Reaction path for the Rearrangement of Methylisocyanide to Isonitrile}
\label{SUPPSEC:RXNPATH}
%======================================
This flowchart, with the appropriate subflowchart in the \texttt{Structure}, \texttt{Reaction Path}, and \texttt{Thermochemistry} steps, was used successfully with a number of \textit{ab initio} and semiempirical Hartree-Fock and \ac{DFT} methods, as reported in the article and shown there in Table 2. These calculations ran smoothly and only had minor problems, mainly due to the different default accuracies of the various methods. Loose convergence criteria for the energy and gradients caused the optimization of the structure of the transition state to fail to converge. The tightening of the convergence criteria for the energy and gradients resolved these issues.

The situation when using several different version of the ReaxFF forcefield\cite{strachan:2003:98301,singh:2013:104114,shan:2014:962} and also different versions of the ANI parameterization of TorchANI\cite{smith:2019:2903,devereux:2020:4192} was quite different. The \ac{NEB} algorithm failed to converge or produced an unreasonable reaction path. For example, in some cases the nitrogen-carbon triple bond of the isonitrile broke, passing through a \ce{CH3N} + \ce{C} atom structure, with the carbon atom then inserting between the methyl group and nitrogen to form isocyanide. In other cases, the methyl-nitrogen bond broke to yield \ce{CH3} and \ce{CN} which then recombined to form the isonitrile. In yet other case, the struture exploeded into separate atoms which then recombined to form the isonitrile. None of these paths is physically reasonable. We also noted that all of the forcefields and machine learning potentials predicted a bent structure for methylisocyanide, suggesting that the potentials had fundamental issues reproducing the isocyanide group.

Unfortunately it is not easy to determine whether the training set for the forcefields or the machine learning potentials contained any isocyanides. Each of the forcefields used in this study were created by modifying existing forcefields to add new species and reactions. This required tracking the history of each forcefield and examining a number of papers and their supplementary information to understand the training set. As far as we can tell, none of the three ReaxFF forcefields examined contained an isocyanide in their training sets, but it is difficult to be certain.

The supplementary material associated with the ANI machine learning potentials contain the training sets used, which were subsets of the structures in available databases such as the GDB \cite{GDB}. Although GDB was produced by an exhaustive generation of plausible structures containing up to, for example, 11 atoms of C, N, O, and F, and hence contains structures with isocyanide functional groups, the selection of a subset does not guarantee that the training set for the ANI potentials contains any isocyanides. The supplementary material, though complete, is only machine readable, so it requires writing a small program to access the data to determine the molecules in the training set and find that there are no isocyanides included. Having access to the training data makes it possible to definitively determine whether specific functional groups, such as isocyanide, were included in the training; however, the need to write a program to access the data is a considerable barrier.

Focusing on the methylisocyanide structure, we can see that all the forcefields and ML potentials examined cannot describe the structure well. As noted above, all methods predict a bent structure rather than the linear structure observed experimentally. In addition, the calculated vibrational frequencies, shown in Table \ref{table:5}

\begin{table}
\centering
\begin{tabular}{*{7}{c}}
\hline\hline
\textbf{Exptl.$^a$} & \textbf{ReaxFF\cite{nguyen:2018:2532}} & \textbf{ReaxFF\cite{shan:2014:962}} & \textbf{ReaxFF\cite{strachan:2003:98301}}  & \textbf{ANI-1x\cite{ANI-1x}} & \textbf{ANI-1cxx\cite{ANI-1ccx}} & \textbf{ANI-2x\cite{ANI-2x}}\\
\hline
263 & 66 & 11 & 68 & 431 & 222 & 258 \\
263 & 288 & 488 & 394 & 981 & 850 & 835 \\
945 & 485 & 1304 & 946 & 1394 & 1304 & 1145 \\
1129 & 1293 & 1333 & 960 & 1532 & 1521 & 1172 \\
1129 & 1294 & 1352 & 1367 & 1757 & 1698 & 1445 \\
1419 & 1949 & 1832 & 1475 & 2019 & 2026 & 1918 \\
1467 & 2075 & 1889 & 1475 & 2031 & 2045 & 1990 \\
1467 & 2075 & 1894 & 2942 & 2155 & 2109 & 2921 \\
2166 & 3309 & 2074 & 3104 & 3241 & 3233 & 4266 \\
2966 & 3388 & 3253 & 3105 & 4258 & 4221 & 4459 \\
3014 & 3388 & 3254 & 3237 & 4354 & 4243 & 4581 \\
3014 & 6952 & 3320 & 9016 & 4391 & 4280 & 4980 \\
\hline\hline
\end{tabular}
\caption{The vibrational frequencies ($\wn$) calculated for methylisocyanide.}
\label{table:5}
  \begin{tablenotes}
    \footnotesize
    \item $^a$ The experimental fundamental frequencies as quoted in Ref.~\citenum{nguyen:2018:2532}.
  \end{tablenotes}
\end{table}

Clearly, some of these frequencies are nonphysical and in general the differences between the calculated and experimental frequencies strongly suggest that neither the ReaxFF forcefields nor the ANI ML potentials can reasonably describe methylisocyanide, let alone the transition state.

%%%%%%%%%%%%%%%%%%%%%%%%%%%%%%%%%%%%%%%%%%%%%%%%%%%%%%%%%%%%%%%%%%%%%
%% Acronyms (alphabetic order)
%%%%%%%%%%%%%%%%%%%%%%%%%%%%%%%%%%%%%%%%%%%%%%%%%%%%%%%%%%%%%%%%%%%%%

\begin{acronym}
    \acro{API}{Application Programming Interface}
    \acro{AQME}{automated quantum mechanical environments for researchers and educators}
    \acro{ASE}{Atomic Simulation Environment}
    \acro{ACF}{autocorrelation function}
    \acro{BSE}{Basis Set Exchange}
    \acro{CI-NEB}{climbing image nudged elastic band}
    \acro{CMS}{computational molecular science}
    \acro{CSV}{comma separated values}
    \acro{DOI}{digital object identifier}
    \acro{DFT}{density functional theory}
    \acro{GUI}{graphical user interface}
    \acro{IDPP}{image dependent pair potential}
    \acro{JSON}{JavaScript Object Notation}
    \acro{LAMMPS}{Large-scale Atomic/Molecular Massively Parallel Simulator}
    \acro{MD}{molecular dynamics}
    \acro{MVC}{model-view-controller}
    \acro{NEB}{nudged elastic band}
    \acro{PyPI}{the Python Package Index}
    \acro{RDBMS}{Relational Database Management System}
    \acro{REST}{Representational State Transfer}
    \acro{RI}{resolution of identity}
    \acro{SEAMM}{Simulation Environment for Atomistic and Molecular Modeling}
    \acro{SDF} {structure-data format}
    \acro{SMILES} {Simplified Molecular Input Line Entry System}
    \acro{SQL} {structured query language}
    \acro{VASP}{the Vienna Ab initio Simulation Package}
\end{acronym}

\bibliography{supplement}